\documentclass{article}
\pdfoutput=1
\usepackage{graphicx}
\usepackage{hyperref}

\usepackage{amssymb,amsthm}
\usepackage{multirow,array}

\newcommand{\p}{\partial}
\newcommand{\n}{\nabla}
\newcommand{\R}{\bf R}
\def\d{\hbox{d}}
\def\eq#1{\begin{eqnarray}#1\end{eqnarray}}
\newtheorem{remark}{Remark}
\newtheorem{proposition}{Proposition}

\title{Continuous and Discrete Adjoints to the Euler Equations for  Fluids
\footnote{\textbf{
Sent for publication to J. Numer. Methods. in Fluid. Mech.
}}}

\author{Fr\'ed\'eric Alauzet\footnote{INRIA, 78153 Le Chesnay, France  ({\tt frederic.alauzet@inria.fr})}
        \and Olivier Pironneau
        \footnote{
        		UPMC-ParisVI,  LJLL, 4 Place Jussieu, Paris, F-75005
        		  ({\tt pironneau@ann.jussieu.fr})}
        }
\begin{document}

\maketitle

\begin{abstract}

Adjoints are used in optimization to speed-up computations, simplify optimality conditions or compute sensitivities.  Because time is reversed in adjoint equations with first order time derivatives, boundary conditions  and transmission conditions through shocks can be difficult to understand. In this article we analyze the adjoint equations that arise in the context of compressible flows governed by the Euler equations of fluid dynamics.  We show that the continuous adjoints and the discrete adjoints computed by automatic differentiation agree numerically; in particular the adjoint is found to be continuous at the shocks and usually discontinuous at contact discontinuities by both.
\end{abstract}

\section*{Introduction}

In optimization adjoints greatly speed-up computations; the technique has been used extensively in CFD  for optimal control problems and shape optimization (see  \cite{jameson,bmop,op,jllcontrol,gunz} and their bibliographies).  Because time is reversed in adjoint equations  and because convection terms operate in the opposite direction, boundary conditions and transmission conditions through shocks can be difficult to understand.  Automatic differentiation in reverse mode \cite{griewank},\cite{tapenade}  automatically generates the adjoint equations so the problem does not arise because boundary conditions are set by the adjoint generator.  But the problem then is to understand and make sure that there is a limit when the mesh size tends to zero and that this limit agrees with the continuous solution.

Answering these questions is important for design of supersonic airplanes because shocks are involved; a good case study is the design of airplaines with the least sonic boom at ground level.  Several investigations have been done already (see B. Mohammadi et al \cite{bmop}, S. Kim et al \cite{alonzo}, A. Loseille et al \cite{alauzet}, M. Nemec \cite{nemec} and the bibliography therein) and automatic differentiation is known to work also in this case. Another application is the optimization of "goal oriented meshes" \cite{rannacher} which minimize the numerical error to compute a functional like drag or lift \cite{alauzet}.

Finite volume schemes for the Euler equations work also for the linearized Euler equations, and so should work equally well on the differentiated Euler system.  However the variables of the differentiated Euler equations are much less regular than those those of the Euler equations: being derivatives of discontinuous functions  the sensitivities of the flow variables have Dirac singularities and it is not at all clear that a finite volume method is capable of computing such solutions; thus the discrete adjoint systems derived by automatic differentiation in reverse mode are based on dubious assumptions.  In other words we may have a convergence problem: does the discrete adjoint state converge to some continuous state? Is it solution of a continuous adjoint equations; and since there may not be a unique way to set up an adjoint for a given optimization problem (see Giles\cite{giles}), which adjoint is it?
\\ \\
All these questions have been addressed in the case of scalar conservation laws in one dimension of space by Ulbrich et al. \cite{ulbrich,gilesulbrich}.  A framework was also introduced in \cite{ulbrich} and \cite{cbop} which can be extended to multidimensional vector systems of conservation laws but the theory of course is lacking.  So we propose here to recall the formalism and do detailed numerical tests to check if the results obtained in one dimension of space extends to two and three dimensions.  Our answers are encouraging: discrete and continuous extended calculus of variations agree; the discrete adjoints seem to converge to the continuous ones; when the mesh is refined the method is stable and gives better solutions.

\section{Preliminaries}
\subsection{Calculus of Variations}\label{CalculusofVariations}

Consider the following problem
\eq{\label{pb1}
	\min_{a}\{ J(u,a)~:~L(u) = f(a)\}
}
where $u\to L(u)$ is  an unbounded operator, like a partial differential system, from a Hilbert space $U$ to another Hilbert space $V$.

 When $a$ varies, everything varies up to higher order terms  according to
\eq{
	\delta J = J'_u\delta u + J'_a\delta a ,\hbox{ with } L'(u)\:\delta u = f'_a\delta a.
}
For Problem (\ref{pb1}), the adjoint state is $u^*\in V^*$, the solution  of  $L^* u^* = J'_u$, i.e.
\eq{\label{defadj}
<L^* \: u^*,v>:=<u^*,L'(u)v> = J'(u)v,~\forall v\in U
}
where $<\cdot,\cdot>$ denotes the duality product between the spaces and their duals, $U$ and $U^*$ in the first instance and $V$ and $V^*$ in the second.

From these definitions we can derive another expression for $\delta J$ because
\eq{	
J'_u\delta u = <L^* \: u^*,\delta u> = <u^*,L' \: \delta u> = <u^*,f'_a \delta a> 
}
  and finally
\eq{\label{dj}
	\delta J = <{f'_a}^*u^* +  J'_a,\delta a>.
}
This is a very useful result because it implies that at the solution we have: ${f'_a}^*u^* +  J'_a=0$, the so-called optimality conditions. It also  says that an iterative scheme like
\eq{
 	a^{m+1}=a^m - \lambda  ({f'_a}^*u^* +  J'_a)^*|_{a=a^m}
}
is a steepest descent algorithm which will converges for a small enough step size $\lambda\in\R^+$.

This calculus is formal but can be justified under continuity and differentiability hypotheses of all functions and operators; 
the difficulty is to show that the higher order terms are indeed small.  

\section{An example with the Euler equations for incompressible flows}

\subsection{The Continuous Case}
Consider an optimization problem for an incompressible inviscid Euler flow. The velocity of the fluid is denoted $u$ the pressure $p$, the domain occupied by the fluid $\Omega$; it could be the 2D cross section of a nozzle with an inflow boundary $\Gamma^-$, an outflow boundary $\Gamma^+$ and walls $\Gamma^0$. One seeks for a nozzle inflow velocity $a$  leading to a velocity $u_d$ in a region of space $D\subset\Omega$:
\eq{\label{pbe}&
\min_a \{J(u)=&\frac12\int_{D\times(0,T)}|u-u_d|^2~:\p_t u +\n\cdot(u\otimes u) +\n p=0,~~\n\cdot u=0,
\cr&&
u_{t=0}=0,~~u|_{\Gamma^-}=a,~~u\cdot n|_{\Gamma^+}=b,~~u\cdot n|_{\Gamma^0}=0
\}
}
where $n$ is the outer normal to $\Gamma:=\p\Omega=\Gamma^-\cup\Gamma^+\cup\Gamma^0$. It is assumed that the integral of $a\cdot n$ on $\Gamma^-$ is equal to the integral of $b$ on $\Gamma^+$.

Calculus of variations is done in  $H_0$,  the space of square integrable divergence free functions in $\Omega$.
\eq{\label{estate}
\delta J = \int_{D\times(0,T)}(u-u_d)\delta u;~~
\int_{\Omega\times(0,T)}v(\p_t\delta u +\n\cdot( \delta u\otimes u + u\otimes\delta u
))=0
}
for all $ v\in H_0$;  $a\otimes b$ is the tensor $a_ib_j$ and below we will use the notation $A:B=\sum_{ij}A_{ij}B_{ij}$.
 Let  $u^*\in H_0^*$ be solution of
\eq{\label{eadj}
\p_t u^*+u\cdot\n u^* - (\n u)u^*+ \n p^* =1_D(u-u_d),~~\n\cdot u^*=0,~~ u^*|_{t=T}=0
}
with $u^*=0$ on $\Gamma^+$, and $u^*\cdot n=0$ on $\Gamma^0\cup \Gamma^-$.

Notice that $\n(u^*\cdot u)=(\n u^*)u+(\n u)u^*$; so by integrating  (\ref{eadj}) multiplied by $\delta u$ and performing an integration by parts :
\eq{&&
\delta J 
 =\int_{\Omega\times(0,T)}\delta u\cdot(\p_t u^* +u\cdot\n u^* +(\n u^*) u + \n(p^*-u^*\cdot u)  )
\cr&&
= - \int_{\Omega\times(0,T)}(u^*\cdot\p_t\delta u+\n \delta u : u\otimes u^* +u ^*\n\cdot(\delta u\otimes u))
\cr&& 
+\int_{\Gamma\times(0,T)}\delta u\cdot(n\cdot u ~ u^* + n\cdot u^* ~ u +(p^*-u^*\cdot u)n)
}
The middle integral reduces to the integral of $-u^*\n\delta p$ because of (\ref{estate}) . Finally, as $u^*\cdot n=0$ on $\Gamma$,
\eq{\label{cgrad}
\delta J = \int_{\Gamma^-\times(0,T)}(a\cdot n ~u^*\cdot\delta a+ (p^*-a\cdot u^*)\delta a\cdot n)
.}

\subsection{The Discrete Case}

Let us use a finite element approximation for $u$ of degree 2 on a triangulation ${\mathcal T}_h$ and of degree 1 for $p$.  With the characteristic-Galerkin method \cite{opCGM} at each time step one has to solve: $\forall v_h\in V_h^2,~~\forall q_h\in V_h^1$, with $v_h|_{\Gamma^-}=0, v_h\cdot n|_{\Gamma}=0$,
\eq{\label{p1p2}&&
	\int_\Omega(\frac{u_h^{m+1}}{\delta t}+\n p^{m+1})v_h - \int_\Omega q_h\n\cdot u_h^{m+1} = \int_\Omega\frac1{\delta t}u_h^m(x-\delta t u_h^m(x))v_h(x)\d x
\cr&&
\hbox{ with } V_h = \{w_h\in C^0(\Omega_h)~:~ w_h|_T\in P^1(T_k)\hbox{ for all triangles }T_k\in{\mathcal T}_h \}
}
It is not hard to prove (see Appendix A) that

\eq{\label{dgrad}
\delta J = \delta t\sum_m\int_\Omega({u_h^*}^m \cdot(u^m_h\cdot\n\delta a_h)+ \delta a_h\cdot({u_h^*}^m\cdot\n u_h^m)
-{p_h^*}^m\n\cdot\delta a_h )
}
with the adjoint state defined by
\eq{\label{incadj}&&
	\int_\Omega(\frac{v_h}{\delta t}+\n q_h){u^*_h}^m - \int_\Omega {p^*_h}^m\n\cdot v_h=\int_D (u_h^{m+1}-u_d) v_h
	\cr&&
+ \frac1{\delta t}\int_{\Omega^+} {{u^*_h}^{m+1}(y+\delta t u_h^{m+1}(x))}v_h(y)\d y 
-\int_\Omega v_h {u_h^*}^{m+1}\cdot \n {u^{m+1}_h}(y) \d x~~~~~
\cr&&
u_h^*(T)=0,~~u^*_h|_{\Gamma^+}=0,~~u^*_h\cdot n|_{\Gamma}=0.  
}
for all $v_h,q_h$ with $v_h|_{\Gamma^+}=0$, $v_h\cdot n|_{\Gamma}=0$.
 In view of the fact that the integral is on the boundaries triangles only and that the gradients are differences from the value of the functions on the boundary and zero on the opposite side of the triangles we see that (\ref{dgrad}) 	is a discretization of  (\ref{cgrad}).
 
 \subsection{Contact Discontinuities: Vortex Sheet}
What happens when the flow is discontinuous as in the case of contact discontinuities?

\begin{figure}
\includegraphics[width=10cm,height=4.5cm]{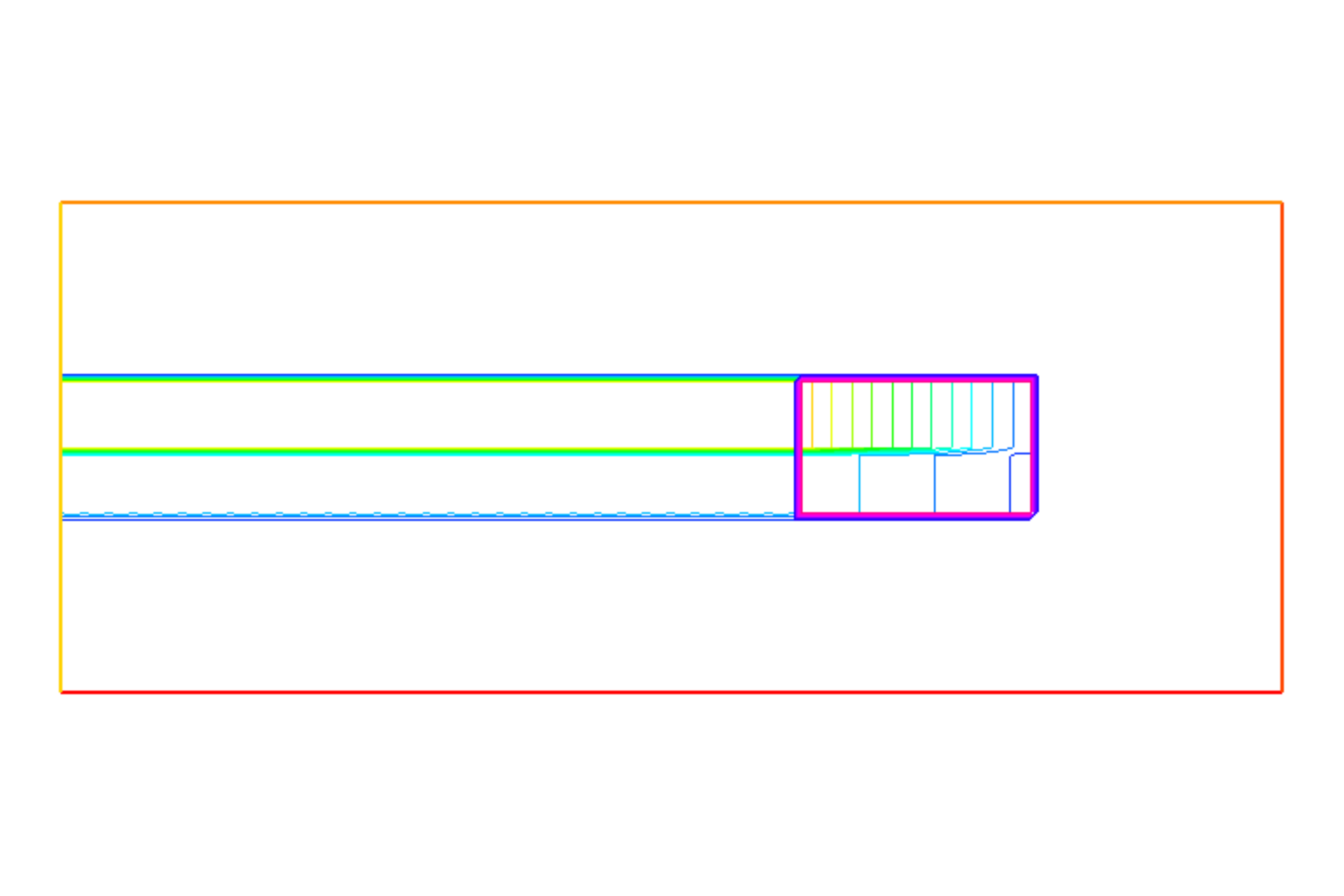}
\includegraphics[width=2cm,height=3cm]{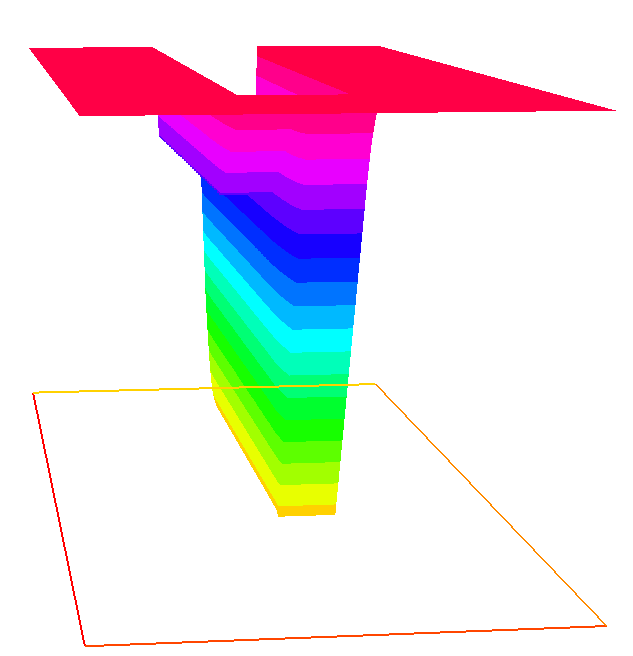}
\caption{\label{figchar}\emph{ Level lines of the adjoint equation to problem (\ref{pbe}). In the bottom part of the channel $u=(2,0)^T$ and in the upper part $u=(1,0)^T$.  The criteria involves an integral of $(u_1-3)^2$ in the box shown in the middle of the channel.  The plot shows the horizontal component of the adjoint state in duality with $u$: it is discontinuous on 3 horizontal lines left of the box and in the box.
The  figure on the right is a front 3D view of the same.}}
\end{figure}

As an illustration we considered $\Omega=(0,7)\times(0,2)$ with $u(0)=[1+1_{y<1},0]^T$, $a(y)=u(0)|_{x=0}$ and other boundary conditions chosen so that $p=0$ on the boundary, with $u_d=(3,0)^T$ and $D=(4,5)\times(0.7,1.3)$.
 Although there is an obvious analytical solution (the solution is stationary : $u=u(0),p=0$) the adjoint has been computed numerically by solving (\ref{incadj}) with the $P^2/P^1$ element; the numerical test shows that  the adjoint is discontinuous (fig \ref{figchar}). The scheme handles the situation perfectly, without oscillation.

\subsection{Conclusion}
From this example we see that both the continuous and the discrete problems can be derived by calculus of variations; however the computations are formal and although the discrete formulae seem to be a discretization of the continuous one it does not appear easy to prove rigorously.

Notice also that by a deriving a discrete adjoint the upwinding term for the adjoint equation is naturally in the right direction.

\section{Generalization to Shocks}
In the presence of shocks, several problems arise with formidable mathematical difficulties: existence and uniqueness of solution, well posedness of the adjoint equations, continuity of these with respect to boundary data, etc.  But there is one difficulty which even a mindless  practitioner cannot ignore: shocks!  Indeed even the formal calculation no longer makes sense in the presence of shocks.

For example if  the density $\rho$ and the velocity $u$ are discontinuous, but with $\rho u$ continuous, it is meaningless to write
\eq{\label{exp}
\delta(\rho u) = u\delta\rho + \rho\delta u
}
because $\delta\rho$ and $\delta u$ contain Dirac singularities which cannot be multiplied by a discontinuous function if both singularities are at the same point.

Indeed let $H(x)$ be the Heaviside function (zero when $x<0$, 1 otherwise); assume that $\rho(x)$ has a shock at $x=x_s(a)$ where $a$ is a parameter:
\eq{\label{heavi}
\rho = \rho^-+(\rho^+-\rho^-)H(x-x_s)
}
Assuming that not only the position of the shock $x_s$ but also $\rho^\pm$ depends on the parameter $a$  then by differentiation
\eq{\label{heavip} &
\delta\rho &= {\rho^-}'\delta a+( {\rho^+}'- {\rho^-}')H(x-x_s)\delta a  - {x_s}'(\rho^+-\rho^-){\underline\delta}(x-x_s)\delta a
\cr&&
=\rho'\delta a -x_s'[\rho]{\underline\delta}(x-x_s)\delta a
}
where primes denote pointwise derivatives with respect to $a$; this is because the derivative of $H$ is the Dirac mass ${\underline\delta}$.  So the Dirac-singularity of $\delta\rho$ is at the same point as the discontinuity of $u$.

\subsection{Extended Calculus of Variations}
Although (\ref{exp}) is not allowed, yet there is a way out:
\begin{equation}\label{prod}
\delta(\rho u) = (\delta\rho){\overline u} + \overline\rho\delta u
\end{equation}
where $\overline\rho$ (and $\overline u$) stands for the mean value of $\rho$ at all points:
\[
\overline\rho(x) = \frac12(\rho(x^+) +\rho(x^-))
\]
This idea  has be used in \cite{ulbrich2,cbop} to extend calculus of variation to discontinuous functions. Indeed when $\rho u$ is continuous across the shock, there is no Dirac mass on the left hand side of (\ref{prod}); let us show that there is no Dirac mass on the right either.

Let $[u]$ denote the jump across the shock, i.e. the  value  on the right side of the shock minus the value on the left side, then on the right hand side of (\ref{prod}) the weight of the Dirac mass is
\eq{&&
-[\rho]x_s'\delta a{\overline u} - [u]x_s'\delta a\overline\rho = -x_s'\frac{\delta a}2((\rho^+-\rho^-)(u^++u^-)+(u^+-u^-)(\rho^++\rho^-))
\cr&&
=-x_s'\delta a[\rho u]=0
}
Even if $[\rho u]\neq 0$ the weights on the Dirac masses are the same on both sides, so (\ref{prod}) is true pointwise and singularitywise at the shock.
\\\\
 More complex functions need to be decomposed into elementary products; for instance to compute the extended variation of $\rho^{-1}$ one writes
\eq{\label{rhom}
1=\frac\rho\rho ~\Rightarrow~0 = \overline\frac1\rho\delta\rho + \bar\rho\delta(\frac1\rho)~\Rightarrow~
\delta(\frac1\rho) = -\overline\frac1\rho\frac1{ \bar\rho}\delta\rho
}
Similarly $\rho^{-1}u^2$ being $(\rho^{-1}) v$ with $v=u u$, the mean value can be computed as such.  For non-rational  functions $f(\rho)$ which cannot be decomposed into elementary products (like $\sin\rho$), the overline operator on derivatives is  the Volpert ratio:

\paragraph{Definition}
\eq{\label{volpert}
\overline{f'(\rho)} :=\left[\matrix{ \frac{[f(\rho)]}{[\rho]}&\hbox{{ at shocks,} }\cr  f'(\rho)& \hbox{{at regular points}}}\right.
}
This is  consistent with (\ref{heavip}) :
\eq{\label{vvv}
\delta f(\rho) &= f'(\rho(x))\delta\rho(x) - [f] x'_s \underline\delta(x-x_s)
&=  f'(\rho)\delta\rho|_{x\neq x_s} - \frac{[f]}{[\rho]}[\rho]x'_s\underline\delta(x-x_s)
\cr&
=  f'(\rho)\delta\rho|_{x\neq x_s}+\overline{f'(\rho)}\delta \rho|_{x=x_s}
&= \overline{f'(\rho)}\delta\rho
}
For vector or matrix valued functions one should use the identity
\eq{
	[F(u)] = \int_0^1 F'_u(s u^+ +(1-s)u^-)\d s
}
and the Volpert ratio on each component.

\section{Shocks and the Adjoint of Burgers' Equation}\label{s4}
\subsection{Analytical check}
First consider the following example:

Given a parameter $a\in\R$, consider the entropy solution of
\eq{ \p_t u + \p_x({u^2\over 2}) = 0 ~~~~u(x,0) = (1+a)(1-H(x)) }
The solution is explicit and its derivative with respect to $a$ can be computed analytically
 \eq{&&
 u=(1+a)(1-H(x-\frac{1+a}2 t)),
 \cr&&
 u_a'=1-H(x-\frac{1+a}2 t)  +\frac{1+a}2 t{\underline\delta}(x-\frac{1+a}2 t)
}
   Let
 \eq{\label{j0}
J(a):={1\over 2}\int_{-\frac12}^\frac12 u(T)^2=\frac14(1+a)^2((1+a) T-1)
}
 Then  $ J'(0) =\frac14(3T-2)$.

To recover this by the extended calculus of variation, one proceeds as follows:
\eq{\label{jj0}
    J'=\int_{-\frac12}^\frac12 \bar u u_a'|_T=\int_{-\frac12}^\frac12 \bar {u^*} u_a'|_0=\int_{-\frac12}^0 {u^*}(x,0)
}
with $u_a'$ solution of
\[
\p_t u_a' +\p_x(\bar u u_a') = 0,~~~ u'_a(0) = 1 - H(x)
\]
 and ${u^*}$ defined by
\eq{\label{badj} \p_t {u^*} + \bar u\p_x {u^*} =0,~ {u^*}|_T= \overline{J'_u}=\bar
u|_T{\bf 1}_{[-\frac12,\frac12]} }
Therefore
\eq{
  {u^*}(x,0)=1-\frac12 H(x+\frac T2)-\frac12 H(x-\frac T2)
}
and (\ref{jj0})  and (\ref{j0}) agree.

Notice that ${u^*}$ is integrated backward in time; it is constant on the characteristics, lines of slope $1+a$ on the left of the shock and of slope $0$ on the right. So the characteristics do not define ${u^*}$ everywhere: there is a "shadow" triangle where no upgoing characteristics cross the line $t=T$. However thanks to the overline on $u$ in (\ref{badj}) there is one characteristic of slope $\frac12(1+a)$ on which ${u^*}=\overline u(T)$, the shock line in fact.
Once ${u^*}$ known on the shock line then any point in the shadow triangle has an upgoing characteristic which crosses the shock and so ${u^*}$ is also known there, constant in this example and equal to $\overline u(T)$.

\emph{Notice that ${u^*}$ is continuous across the shock and discontinuous at the characteristics that meet the shock at $t=T$.}

\subsection{Numerical check}
Whether a numerical scheme will be capable of such a reconstruction is a matter of wonder.

Consider another example with $T=2$ for which one seeks to compute the sensitivity with respect to $a$ at $a=0$:
\eq{\label{simple}
	J(a)=\frac12\int_{\R^+} |u(x,T)|^2\d x~:~
	\p_t u + \p_x\frac{u^2}2 =0,~ u|_0=-\min(\hbox{atan}(x+a),0)
}
Although there is no shock at time zero, one develops later.
The following conservative finite difference scheme has be used
\eq{\label{fdm}&&	
\frac{u^{m+1}_i-u^m_i}{\delta t} +
	s_i\frac{{u^m_i}^2-{u^m_{i-1}}^2}{2\delta x} +(1-s_{i+1})\frac{{u^m_{i+1}}^2-{u^m_{i}}^2}{2\delta x}=0
	\cr&&
	\hbox{with } s_i=1 \hbox{ if } u^m_i+u^m_{i- 1} >0,~\hbox{ 0 otherwise}~~~
}
When $J$ is approximated by $\frac12\sum_{i:x_i>0}|u_i^M|^2$,  the discrete Adjoint to (\ref{fdm}) is,
\eq{\label{disad}&&
\frac{{u^*}^{m-1}_i-{u^*}^m_i}{\delta t} +\frac{ u^m_i}{\delta x}(
	{u^*}^m_i s_i - {u^*}^m_{i+1}s_{i+1} +{u^*}^m_{i-1} (1-s_i) - {u^*}^m_{i}(1-s_{i+1}))=0
\cr&&
{u^*}^M_i= u_i^M
}
with $s_i =s_i^+-s_i^-=\hbox{sign}(u^m_i+u^m_{i-1})$. Then 
\[
\delta J = -\sum_{i=1}^{N-1}(-{u^*}^0_i+\frac{\delta t}{\delta x}u^0_i(
	{u^*}^0_i s_i - {u^*}^0_{i+1}s_{i+1} +{u^*}^0_{i-1} (1-s_i) - {u^*}^0_{i}(1-s_{i+1}))\delta u^0_i\delta x
\]

\begin{figure}[htdp]
\begin{center}
\includegraphics[width=6cm]{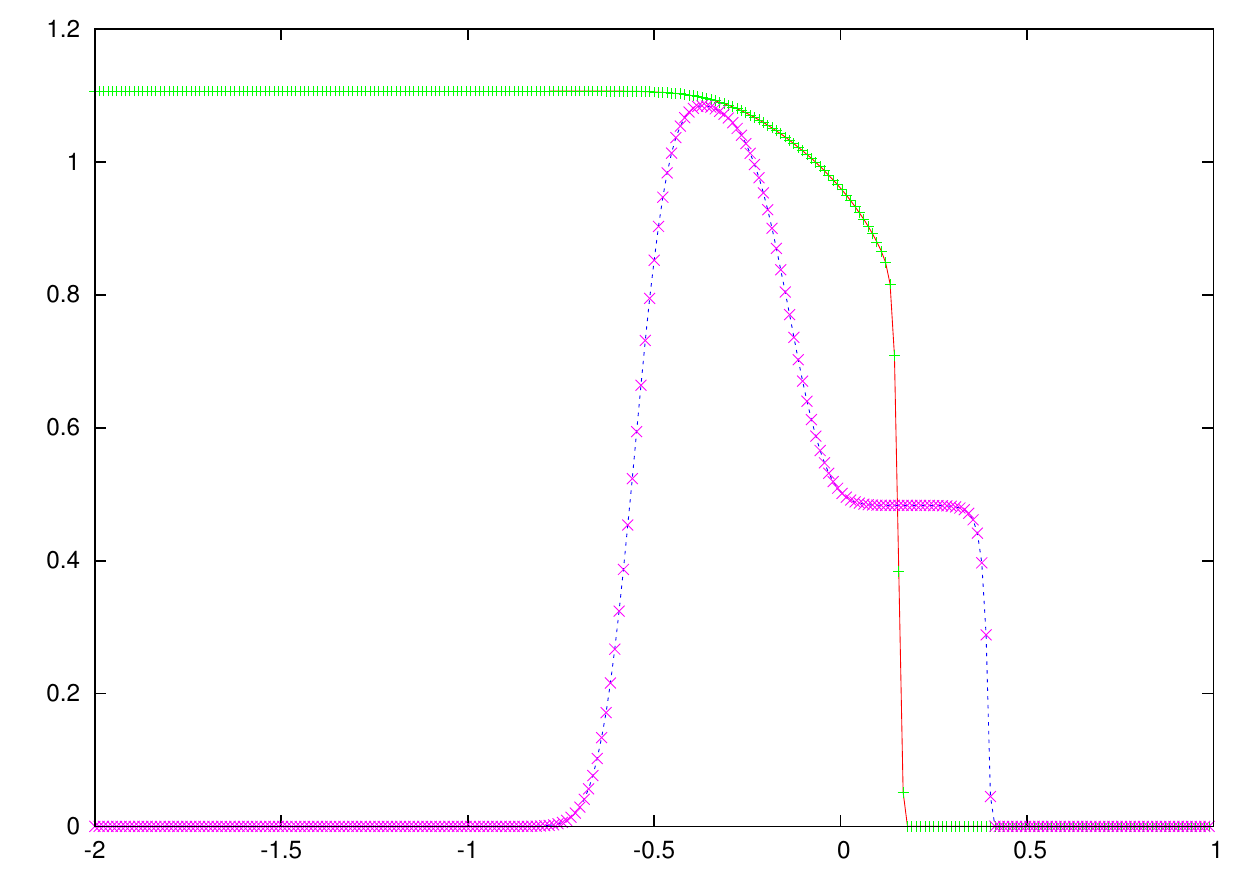}
\includegraphics[width=6cm]{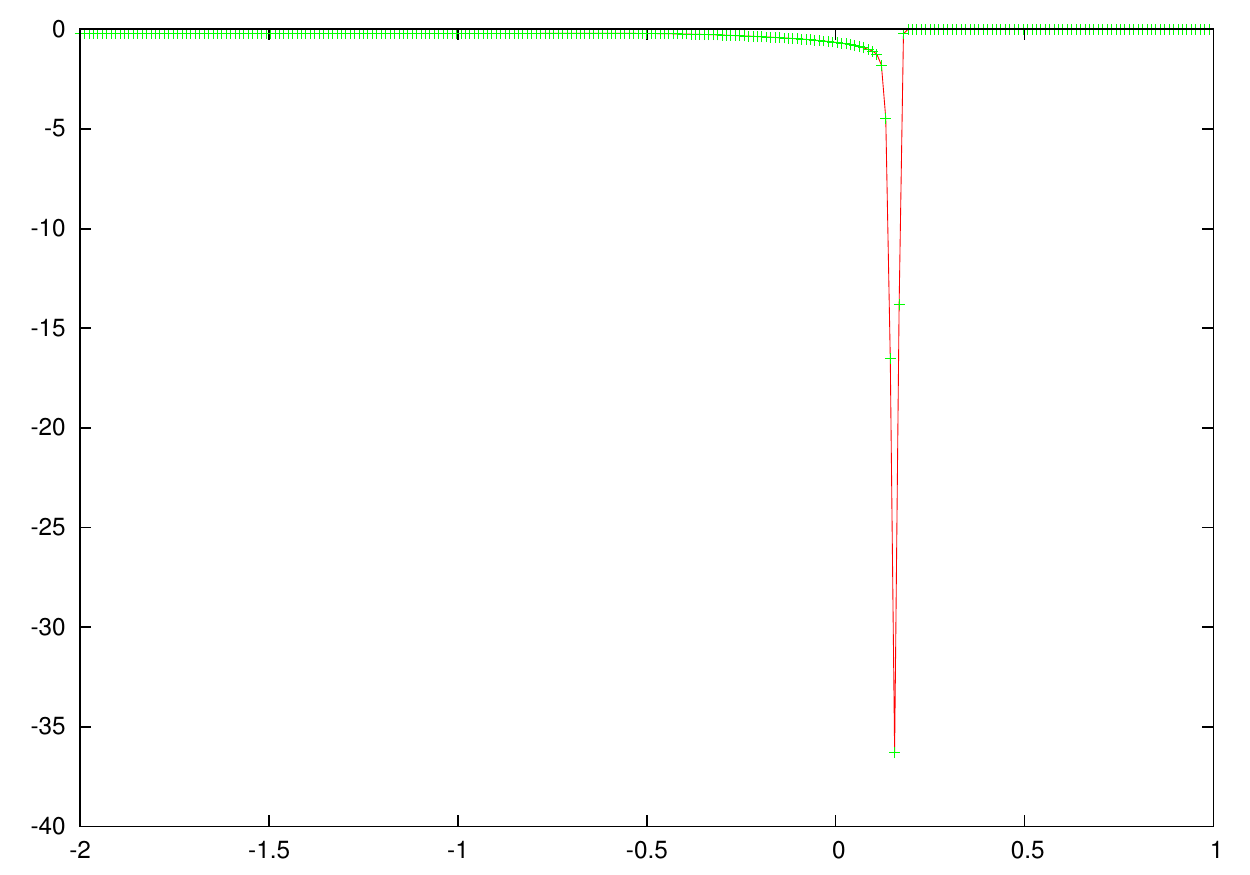}\\
\caption {Left displays at T=0.325 $u$ (green) solution of Burgers' equation and the adjoint state $u^*$ (red), while $u'_a$  is shown on the right.  Notice the Dirac-like singularity of $u'_a$.}
\label{figtwo}
\end{center}
\end{figure}

On this simple example the adjoint ${u^*}$ seems to be correctly calculated by the conservative scheme (see Figure \ref{figtwo}).

The results of Table \ref{tableone} show that the scheme computes $J$ correctly. Automatic differentiation of the computer program also gives the correct  $J_a'(0)$. On  figure \ref{figtwo} the solution and its derivative with respect to $a$ are shown.
\begin{table}[htdp]
\begin{center}
\begin{tabular}{|c|c|c|c|c|}
\hline
$J(0)$ & $ J(\delta a)$  & $(J(\delta a)-J(0))/(\delta a)$& $ J^\prime_a(0)$ &  $^A J^\prime_a(0)$   \cr
\hline
0.195009 &  0.190057  & -0.4952&  -0.492863&  -0.492863\cr
\hline
\end{tabular}
\end{center}
\caption{ Scheme (\ref{fdm}) gives the correct result for $J$ and $J^\prime_a$. 
A computation of $J^\prime_a(0)$ has also been done by finite difference with $\delta a=0.01$ then  with the discrete gradient (\ref{disad}) using the discrete adjoint and finally with Automatic Differentiation : $^A J^\prime_a(0)$.
\label{tableone}}
\end{table}%

\subsection{Stationary problems}
 The stationary case is difficult but it can be investigated as a limit case of the time dependent problem. As an example  consider
\[
J:={1\over 2T}\int_{R\times(0,T)}u^2~:~ \p_t u + \p_x({u^2\over 2}) = 0 ~~~~u(x,0) = (1+a)(1-H(x) )
\]
Then
\[
    J'={1\over T}\int_{R\times(0,T)}\overline{u}u'
~~\hbox{with}~
\p_t u' + \p_x(\bar u u') = 0 ~~~~u'(x,0) = (1-H(x))
\]
The adjoint equation is
\eq{\p_t u^* + \bar u\p_x u^* =\frac1T \overline{u},~ u^*|_T= 0
}
and then at $a=0$
\[
J'=\int_R u'(0)u^*(0)=\int_R(1-H(x))u^*(0) =\int_{-\infty}^0u^*(0)
\]
Here too $u^*$ is continuous at the shock.

\subsection{Convergence}

For scalar conservation laws in 1D the extended calculus of variation has been justified by S. Ulbrich \cite{ulbrich}. He introduced the concept of shift derivatives for functions of bounded variations on an interval $I=[a,b]$.  The idea is that if $\rho\in BV(I)$ it may have discontinuities at $a\in I$, so - as explained in (\ref{heavi},\ref{heavip}) - on top of the usual variation in magnitude $\delta\rho$, the variations of the position of the discontinuities should also be considered. The result is not a function but a distribution, however a Dirac mass can be approximated by a large function on a small support, so (\ref{heavip}) can also be written as
\eq{\label{heavipp}
\delta\rho = {\rho}'_a\delta a - {x_s}'_a[\rho](H(x-x_s-\delta a)-H(x-x_s))
}
Hence  Ulbrich suggests to focus on what he calls \emph{shift-variations} of $\rho$ which includes at singular points $a:=x_s$ of $\rho$
\eq{&&
  \delta_a\rho  :=  \hbox{sign}(\delta a)[\rho(a)]^-{\bf 1}_{(a-\delta a^-,a+\delta a^+)}
}	
where $[\rho]^-(x)=-\min(0,\rho(x^+)-\rho(x^-))$ denotes the negative part of the jump and $\delta a^\pm$ the positive and negative part of $\delta a$.  Thus a general shift-variation of $\rho$ is
\eq{&&
 \tilde\delta\rho  :=  \delta\rho(x) + \sum_k\delta_{a_k}\rho
}	
Then $u\to y(u)$ is said to be shift-differentiable at $u$ if there exists a linear operator $y'_u$ and a finite set of values $y'_k$ associated with a set of singularity points $a_k$  such that
\eq{
y(u+\delta u) = y(u) +y'_u(u)\cdot\delta u +\sum_k y'_k\delta_{a_k}u + o(\|\tilde\delta u\|)
  }
With this definition discontinuous functions can be expanded by a Taylor formula but only for a restricted class of $\delta u$.

Furthermore when it exists the shift derivative is also the distribution derivative such as the one computed for (\ref{heavi}) above.

Ulbrich\cite{ulbrich} proves that entropy solutions of scalar conservation laws in 1D are shift differentiable with respect to initial data under very reasonable hypotheses; the framework gives a natural way to define an adjoint and the duality formula (\ref{defadj}) necessary for the extended calculus of variation to hold.  In an other article Ulbrich and Giles \cite{gilesulbrich} have studied the convergence of the discrete adjoint to the continuous one in a simple case similar to (\ref{simple}).

\section{Extension to the Full Euler System in 2D}
Let us apply the same formalism to the two dimensional Euler equations for fluids:
\begin{equation}\label{euler2d}
\displaystyle{\partial_t W} + \nabla \cdot F(W) = 0 \,, W(0)=0
\end{equation}
where $W=(\rho,\rho u,\rho v, \rho E)^T$ is the vector of conservative variables
and the tensor $F$ represents the convective flux: $F(W) = F_1(W) \,e_x + F_2(W) \,e_y$ with
$$
F_1(W) = \left( \begin{array}{c}
\rho u \\ \rho u^2 + p \\ \rho u v \\ (\rho E + p) u
\end{array} \right) \; \mbox{and} \; \;
F_2(W) = \left( \begin{array}{c}
\rho v \\ \rho u v \\ \rho v^2 + p \\  (\rho E + p) v
\end{array} \right)
$$
Additionally we assume that $W$ is given at time $t=0$, and for bounded domains $\Omega$ we add  boundary conditions which we will discuss later at length and which we denote loosely by $W\in W_{ad}$.

\bigskip

We wish to measure the change $\delta J$ of a functional $J$ when something changes in the data (parameters) of the above problem. 

For the silent business jet problem the functional is an integral of  the pressure on a part $S$ of the boundary of the domain (the ground);  for the control of the outflow of a scramjet $J$  is an integral of the density on the outflow boundary. In both cases there exists a linear map  $W\to B(W)\in \mathbb R^d$ and a vector $b\in \mathbb R^d$ such that
\begin{equation}\label{J}
J = \frac12\int_{S\times(0,T)}|B(W)-b|^2
\end{equation}
The parameter (denoted by $a$ in the previous section) could  be  in the shape of the business jet or of the scramjet but at this level it is not important. The extended calculus of variation on $J$ gives :
\begin{equation}\label{J1}
\delta J = \int_{S\times(0,T)}\overline{(B(W)-b)B'}\delta W,
 \end{equation}
Turning to $\delta W$ we know from above that it satisfies
\begin{equation}\label{deltaeuler}
\displaystyle{\p_t \delta W} + \nabla \cdot (\overline{F'(W)}\delta W )= 0 \,, \delta W(0)=0,~~~\delta W-\underline A\delta a\in W_0
\end{equation}
where $W_0$ is the tangent space to $W_{ad}$ at $W$ and $\underline A\delta a$ is the (possible) change of boundary condition du to the change $a\to a+\delta a$.
 So let us introduce the adjoint equation
 \begin{equation}\label{adjS}
\displaystyle {\partial_t  W^*}+ \overline{F'(W)}^T\nabla  W^*
= 0,~~~  \; W^*(T)=0. 
\end{equation}
with boundary conditionsto be chosen later.
In variational form (\ref{adjS}) is
 \begin{equation}\label{deltaeulervar}
\displaystyle \int_{\Omega\times(0,T)}({\partial_t V}\cdot  W^* -\overline{F'(W)}^T\nabla  W^*
\cdot V)+ \int_\Omega W^*(0)\cdot V(0)=0~~~~\;\forall V.
\end{equation}
 Let us choose $V=\delta W$, integrate by parts and use (\ref{deltaeuler}):
\begin{eqnarray*}&&
0= \displaystyle \int_{\Omega\times(0,T)}({\partial_t\delta W}
\cdot  W^* -\overline{F'(W)}^T\nabla  W^*\cdot\delta W)
\cr &&=
\int_{\Omega\times(0,T)} W^*\cdot(\displaystyle{\partial_t \delta W}+ \nabla\cdot(\overline{F'(W)} \delta W))
- \int_{\partial\Omega\times(0,T)}W^*\cdot ( n\cdot(\overline{F'(W)} \delta W)
\end{eqnarray*}
Hence
 \begin{equation}\label{dual}
\int_{\partial\Omega\times(0,T)}W^*\cdot ( n\cdot(\overline{F'(W)} \delta W)=0
\end{equation}
This gives a method to evaluate $\delta J$ but for each case the boundary conditions for $W^*$ are different because they must be chosen so that  $\delta J$ appears in (\ref{dual}) .

\subsection{An example: the scramjet}

Consider a scramjet (see figure \ref{fig6}) with supersonic flow at the inlet and the outlet on which we observe the density at the outlet $S$ and aim at tuning the parameters so that it be constant.

When the flow is stationary and $J$ is the integral on $S\times(0,T)$ of $(\rho/\rho_\infty - 1)^2/(2T)$  then, asymptotically in $T$, $B=(1,0,0,0)/\rho_\infty^2$ and $b=1$.

When the flow is hypersonic there is no condition on $W$ at the outflow boundary $S$, therefore $W_{ad}$ is the whole space. Let $W^*$ on $S$ be such that
\eq{\label{50}&&
{W^*}^T n\cdot(\overline{F'(W)}=\overline{(B W -b)B} \hbox{ on } S
}
With $q^2=u^2+v^2$, $S$ being parallel to the $y$-axis, this spells out as
{\small
\begin{eqnarray}\label{50a}&&
{W^*}^T
\left( \overline{\begin{array}{cccc}
0 &1 &0& 0 \\
\displaystyle \frac{(\gamma-1)}{2} q^2  - u ^2 &  (3-\gamma)u &   (1-\gamma) v & (\gamma -1)  \\
\displaystyle  - v u & v & u &0\\
\displaystyle \left( (\gamma -1) q^2 - \gamma E \right) u &
\displaystyle \left( \frac{p}{\rho} + E \right)  - (\gamma-1) u^2&
\displaystyle  (1-\gamma) v u & \gamma u
\end{array} }\right) ~~~~
\cr&&
~~~~~~~~~~~~~~~~~
=\frac1\rho_\infty(\frac{\bar\rho}\rho_\infty - 1)\left(\matrix{1& 0& 0 & 0}\right)
\end{eqnarray}
}
It is a non singular linear system for $W^*$ on $ S$ solvable as
\eq{\label{adjcond}
W_4^* &=& \frac{\displaystyle \frac1\rho_\infty(\frac{\overline \rho}\rho_\infty - 1)}
{\displaystyle \left( \frac{\gamma-2}{2} \overline q^2 + \frac{\gamma}{\gamma-1} \overline u^2 + \overline v^2 - \gamma \overline E  \right) \, \overline u} \\
W_3^* &=& -\overline v \, W_4^* \\
W_2^* &=& -\frac\gamma{\gamma-1} \overline u\,  W_4^* \\
\label{adjcond1}
W_1^* &=& \left(\frac{\gamma+1}{\gamma-1} \overline u^2 + \overline v^2- \frac {\overline p}{\overline \rho} -\overline E \right) W_4^*
}

\section{Numerical Flow Solvers}\label{WOLF}
In this section we wish to 
Implement numerically (\ref{adjcond})-(\ref{adjcond1}), and compare with the conditions given by Automatic Differentiation of the Euler solver.

\subsection{Euler Flow Solvers}
\begin{figure}[!b]
\centering
\begin{tabular}{c}
\includegraphics[height=5cm]{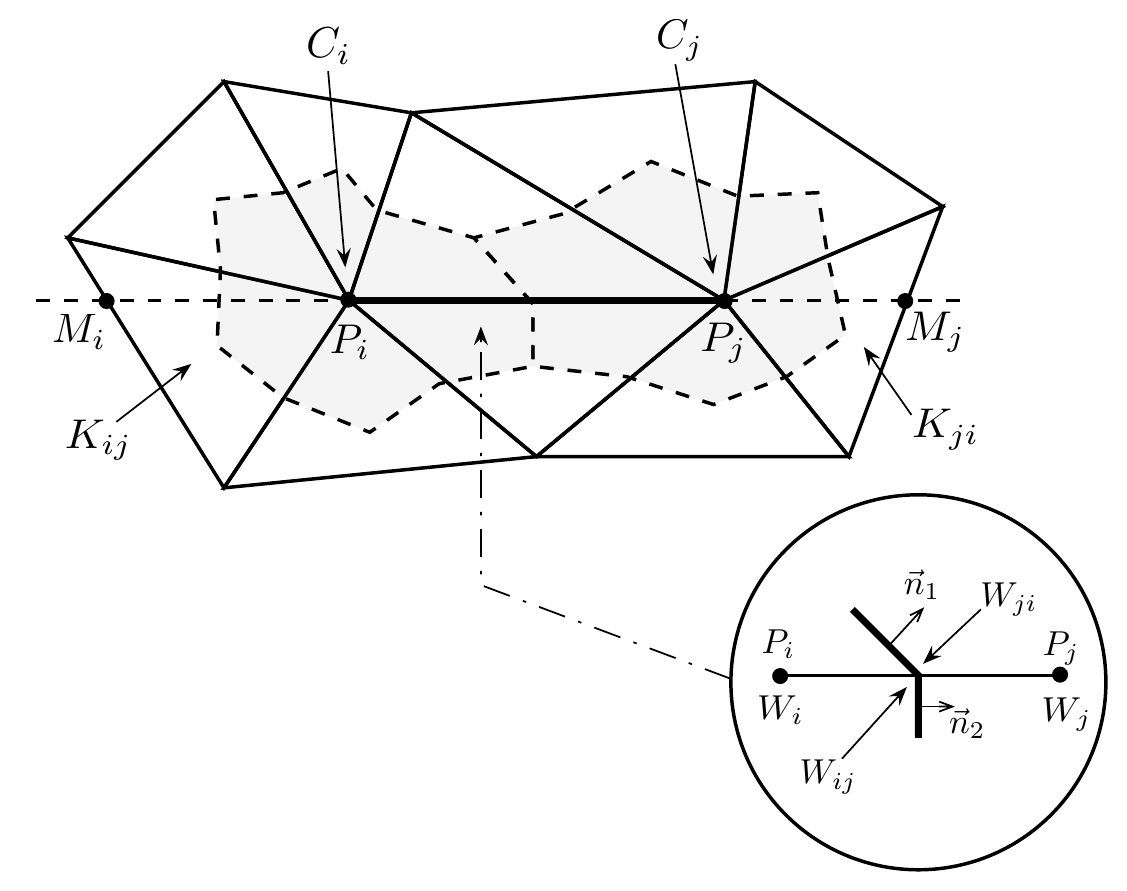}
\end{tabular}
\caption{\label{muscl} \it Two finite volume cells $C_i$ and $C_j$, and the upwind triangles $K_{ij}$ and $K_{ji}$ associated with edge $P_iP_j$. The common boundary $\partial C_{ij}$ with the representation 
of the solution extrapolated  for the MUSCL correction is shown.}
\end{figure}

The proposed method is a vertex-centered finite volume scheme applied to simplicial unstructured meshes 
which uses a particular edge-based formulation with upwind elements~\cite{Alauzet-2010b,Debiez-2000}. 
The vertex-centered finite volume formulation consists in associating with each vertex $P_i$ of the mesh a 
finite volume cell, denoted $C_i$, which is built by joining each vertex to the middle point of its opposite edge. 
The common boundary between two neighboring cells $C_i$ and $C_j$ is called $\partial C_{ij}=\partial C_i \cap \partial C_j$. An illustration of this construction in two dimensions is shown in Figure~\ref{muscl}.

The finite volume method assumes $W$ piecewise constant on the cells and integrates (\ref{euler2d}) on each cell $C_i$:
\begin{equation}\label{formul_variationnelle2}
|C_i| \frac{dW_i}{d t} + \int_{\partial C_i}F(W_i) \cdot {\bf n}_i \,d\gamma 
= |C_i| \frac{dW_i}{d t} + \sum_{P_j \in {\mathcal V}(P_i)} F|_{{ij}} \cdot \int_{\partial C_{ij}} {\bf n}_i \,d\gamma = 0 \,,
\end{equation}
where ${\bf n}_i$ is the outer normal to cell $C_i$,  
${\mathcal V}(P_i)$ is the set of all neighboring vertices of $P_i$
and $F|_{{ij}}$ represents the constant value of $F(W)$ at interface $\partial C_{ij}$ from the $P_i$-side. 
The flow is calculated with a numerical flux function, denoted $\Phi_{ij}$:
\begin{equation}\label{flux_numerique}
\Phi_{ij}  := \Phi_{ij}(W_i,W_j,{\bf n}_{ij}) = F|_{{ij}}  \cdot \int_{\partial C_{ij}} {\bf n}_i \,d\gamma \,,
\end{equation}
where $\displaystyle {\bf n}_{ij} = \int_{\partial C_{ij}} {\bf n}_i \,d\gamma$. 
Several upwind numerical flux functions are available and can be formally written:
\begin{equation}\label{fluxnum}
\Phi_{ij}(W_i,W_j,{\bf n}_{ij}) = \frac{F(W_i)+F(W_j)}{2} \cdot {\bf n}_{ij} + d\left(W_i,W_j,{\bf n}_{ij}\right) \,,
\end{equation}
where the function $d\left(W_i,W_j,{\bf n}_{ij}\right)$ contains the upwind terms and depends 
on the chosen scheme. 
Here the Roe approximate Riemann solver described in \cite{Roe-1981} is used.

\paragraph{Roe's approximate Riemann solver} is based on the Jacobian  $F'(W)$ :
\eq{\label{roeflux}
\Phi^{Roe}(W_i,W_j,{\bf n}_{ij}) = \frac{F(W_i)+F(W_j)}{2}
 \cdot {\bf n}_{ij} + |\tilde A(W_i,W_j)| \frac{W_i-W_j}{2} 
}
where $\tilde A=F'(\tilde W)$ evaluated for the Roe average variables 
$\tilde W:=(\tilde \rho, \tilde {\rho u}, \tilde{\rho v}, \tilde{\rho E})$  as explained in \cite{Roe-1981}.
For a diagonalizable matrix $A = P \Lambda P^{-1}$, $|A|$ stands for $|A| = P |\Lambda| P^{-1}$.
The eigenvalues of $F'(W)$ are real and equal to $u$, $u+c$ and $u-c$. 

\paragraph{Higher-order Finite Volume Methods.}
The previous formulation gives at best only a first-order scheme,  
higher-order extensions are possible with a MUSCL addition.  
The idea is to use extrapolated values $W_{ij}$ and $W_{ji}$ of $W$ at the interface $\partial C_{ij}$ 
to evaluate the flux, cf. Figure~\ref{muscl}. The following approximation is performed:
$$
\Phi_{ij} = \Phi_{ij}(W_{ij},W_{ji}, {\bf n}_{ij}) \,, 
$$
with $W_{ij}$ and $W_{ji}$ which are linearly extrapolated by:   
\begin{equation}\label{interpolmuscl}
W_{kl} = W_k+\displaystyle \frac{1}{2} \, (\nabla W)_{kl} \cdot \overrightarrow{P_kP_l} 
\quad \mbox{ with }kl=ij or ji,
\end{equation}
where the approximate ``slopes" 
$(\nabla W)_{ij}$ and $(\nabla W)_{ji}$ are defined on the edges by using a 
combination of centered, upwind and nodal gradients~\cite{Alauzet-2010b}. 

Such MUSCL schemes are not monotone. Therefore, limiting functions must be coupled 
with the previous high-order gradient evaluations to make sure that the scheme is TVD. 
Then $\n W_{kl}$ in (\ref{interpolmuscl}) is substituted 
by a limited gradient denoted $(\nabla W)_{ij}^{lim}$. 
In this work we have used the three-entries limiter introduced by Dervieux which is a generalization 
of the Superbee limiter~\cite{Cournede-2006}.

\paragraph{Boundary conditions.}

\emph{Slip boundary conditions} are imposed for the walls because the flow is inviscid: ${\bf U} . {\bf n} = 0$. 
This boundary condition is imposed weakly by setting:
\begin{equation}\label{fluxslipping}
\Phi^{Slip}(W_i) = (0,p_i\,{\bf n}_i,0)^t \,.
\end{equation}

\emph{For free-stream external flow conditions}
at inflow boundaries  $\Gamma_\infty$ which approximate infinity, a free-stream uniform flow $W_\infty$ is known. 
There,  the Roe flux is replaced by the Steger-Warming flux~\cite{Steger-1981}:
\begin{equation}\label{fluxstream}
\Phi^\infty(W_i) = A^+(W_i,{\bf n}_i) W_i + A^-(W_i,{\bf n}_i) W_\infty \,,
\end{equation}
where $\displaystyle A^\pm = \frac12(|A|\pm A)$ and $A=F'(W)$. 

\emph{At free-stream outflow boundaries} no condition is imposed so nothing is done to the finite volume scheme.

\paragraph{Time discretization.}
Once the equations have been discretized in space, 
a set of ordinary differential equations in time is obtained: $W_t - L(W) = 0$.
To discretize in time, a high-order multi-step Runge-Kutta scheme is considered. 
Such time discretization methods, called SSP (Strong-Stability-Preserving), have non-linear stability 
properties which are particularly suitable for the integration of system of hyperbolic conservation laws 
where discontinuities appear. 
The optimal $2$-stage order-$2$ SSP Runge-Kutta scheme introduced by Shu and Osher~\cite{Shu-1988} 
is used in this study: 
\begin{eqnarray*}
W^{(1)} & = & W^n + \delta t \, L(W^n)  \,, \\
W^{n+1} & = & \frac{1}{2} W^n + \frac{1}{2} W^{(1)} + \frac{1}{2}  \delta t \, L(W^{(1)})    \,,
\end{eqnarray*}
The scheme is stable with a $CFL$ coefficient close to $1$.
Alternatively for ~(\ref{euler2d}) an implicit time scheme can be used, such as Implicit Euler:
\eq{\label{implicitEuler}
|C_i| \frac{W_i^{n+1}-W_i^n}{\delta t^n_i} + {\Psi_{i}^{2}}W^{n+1}=0 \,,
}
where ${\Psi_{i}^{2}}(W^n)$ is the second order numerical flux associated with cell $C_i$ given by the sum of the 
volume fluxes~(\ref{fluxnum}) and boundary  fluxes. 

\section{The Numerical Scheme for the Adjoint}

The adjoint of the discrete system is easier to establish when (\ref{implicitEuler}) is used. 
The linearized version of (\ref{implicitEuler})  is
$$
\left( \frac{|C_i| }{\delta t^n_i} I_d +  {\Psi_{i}^{2}}'(W^{n}) \right) \left( W_i^{n+1}-W_i^n \right)
= - {\Psi_{i}^{2}}(W^{n}) \,.
$$
As the second order flux is too difficult to differentiate by hand, the second order Jacobian matrix is approximated by the first order one leading to:
$$
\left( \frac{|C_i| }{\delta t^n_i} I_d + {\Psi_{i}^{1}}'(W^{n}) \right) \left( W_i^{n+1}-W_i^n \right)
= - {\Psi_{i}^{2}}(W^{n})
$$
Thus the variation of the state variables $\delta W$ is given by
$$
{\mathcal A}^n \, \delta W^{n+1} = - \Psi^{2}(W^{n}) \,.
$$
Consequently, following Section~\ref{CalculusofVariations}, for Problem~(\ref{euler2d}), for a given functional $J(W)$, the adjoint state $W^*$   
 is defined by 
\eq{\label{approxadj}
{\mathcal A^n}^* \, W^* =J'(W) 
}
where ${\mathcal A^n}^*$ is the transposed of matrix ${\mathcal A^n}$. Note however that it suffices to take
${\mathcal A^n}^*=({\Psi_{i}^{1}}'(W^{n}))^T$
because at convergence the solution does not depend on time and $\Psi_{2}(W^{n})$ tends to 0.

\subsection{Automatic differentiation of the numerical flux}

Tapenade \cite{tapenade} is a software which takes fortran or C programs as input and returns the differentiated programs; 
it is done by copy-paste of the program on the web page of the Tapenade internet site.  
One must specify the variable(s) with respect to which the differentiation is done. 
It works in direct or reverse mode and the later is very similar to the adjoint method.  
However the result may not be optimal. For instance if a program to compute the solution of $F(u,a)=0$ 
uses an iterative scheme like Newton's the differentiated code produced by Tapenade will also use 
Newton's method to compute $u'_a$ such that $F'_a(u,a)+F'_u(u,a)u'_a=0$ even though this 
is a linear system which can be solved directly. 

In this study  everything was differentiated by hand except the C-function which implements the numerical flux function which was passed to Tapenade for differentiation in direct mode.  This way the full flux functions with MUSCL correction and slope limiter were differentiated.  Alternatively we also wrote a solver where only a portion of the flux function is differentiated as explained below.  Although full differentiation is safer,  as we shall see, numerical test
showed no difference when incomplete differentiation is used.

\subsection{Manual approximate differentiation of the numerical fluxes}

One reason to prefer differentiation by hand of the flux functions of the finite volume method is complete control over the source code but also execution speed.

\paragraph{Differentiation of Roe's flux}
In the Roe flux presented in Section~\ref{WOLF} and given by (\ref{roeflux}) the Roe matrix ({i.e., the dissipative matrix) is not differentiated:
\begin{eqnarray*}\label{approxAD}
{\Phi'_{W_i}}^{Roe}(W_i,W_j, {\bf n_{ij}} ) 
& \approx & \frac{1}{2} \left( F'(W_i)\cdot {\bf n_{ij}}  + |\tilde A(W_i,W_j)| \right)  \\
& = & \frac{1}{2} \left( A(W_i) + |\tilde A(W_i,W_j)| \right)  \,,
\end{eqnarray*}
and similarly for ${\Phi'_{W_j}}^{Roe}(W_i,W_j, {\bf n_{ij}} ) $.
The analytic expression of $A(W)$ is given in Appendix B.

\subsection{Differentiation of the boundary conditions.}
In order to differentiate the boundary conditions, we linearize the integrals which
involve a boundary flux. Indeed, a Dirichlet condition does not require any computation 
as it fixes some state variable at a given value. 

The flux associated with the slip boundary condition is given by  (\ref{fluxslipping}), hence using (\ref{JacP}) in Appendix B,
the following Jacobian is obtained:
$$
{\mathcal A_i^{Slip}} = (\gamma-1) \left( \begin{array}{cccc}
0 & 0 & 0 & 0 \\
\displaystyle \frac{q^2}{2}n_x & -u n_x & -v n_x & n_x \\ 
\displaystyle \frac{q^2}{2}n_y & -u n_y & -v n_y & n_y \\ 
0 & 0 & 0 & 0 
\end{array} \right) 
\quad \mbox{where} ~~ q^2 = u^2+v^2. 
$$
For the Steger and Warming type boundary conditions, where the flux is given by (\ref{fluxstream}), 
the linearization is direct: 
$$
{\mathcal A_i^\infty} = A^+(W_i, {\bf n}_i) \,.
$$
\section{Numerical Results for a Scramjet with Outflow Control}
\label{sec31}
A scramjet with supersonic flow at the inlet and the outlet has been computed with the flow solver described previously.  The discrete adjoint (\ref{approxadj}) was generated  by hand using exact differentiated flux generated by Tapenade or manually  using (\ref{approxAD}).

When the flow is stationary and $J$ is the integral on $S\times(0,T)$ of $(\rho/\rho_\infty - 1)^2/(2T)$ where $S$ is the outflow boundary (parallel to the $y$ axis), then we have shown analytically that the boundary conditions at the outflow for the adjoint $W^*$ are given by (\ref{adjcond}).

\subsection{Numerical Results}
The flow and the adjoint are shown on figure \ref{fig6}
\begin{figure}
\begin{center}
\includegraphics[width=12cm]{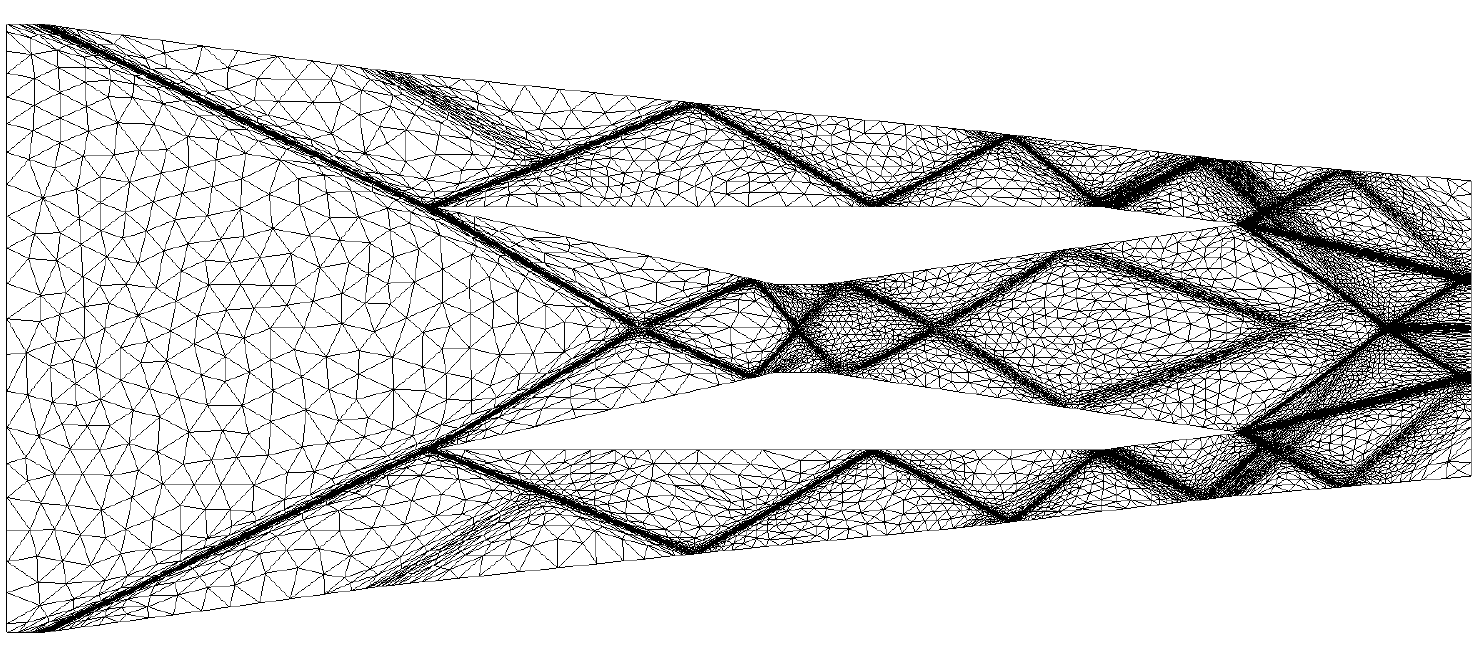}
\includegraphics[width=12cm]{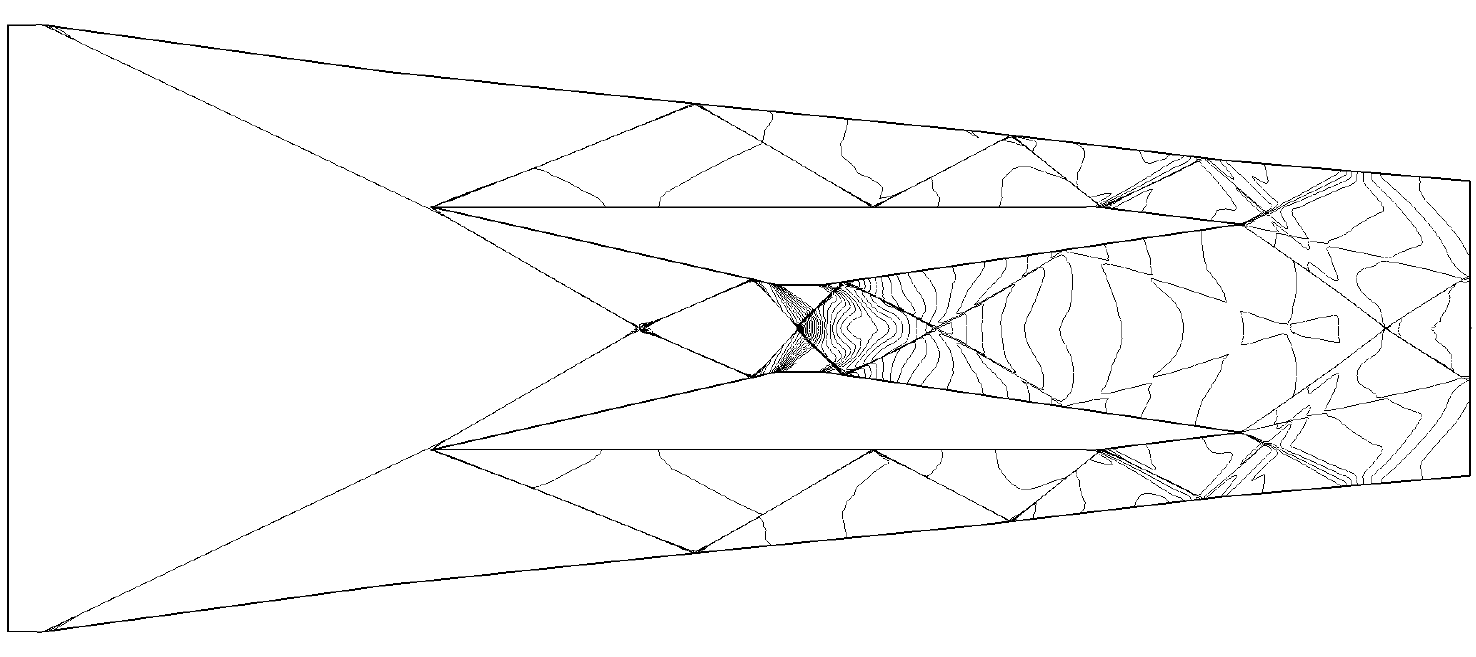}
\includegraphics[width=12cm]{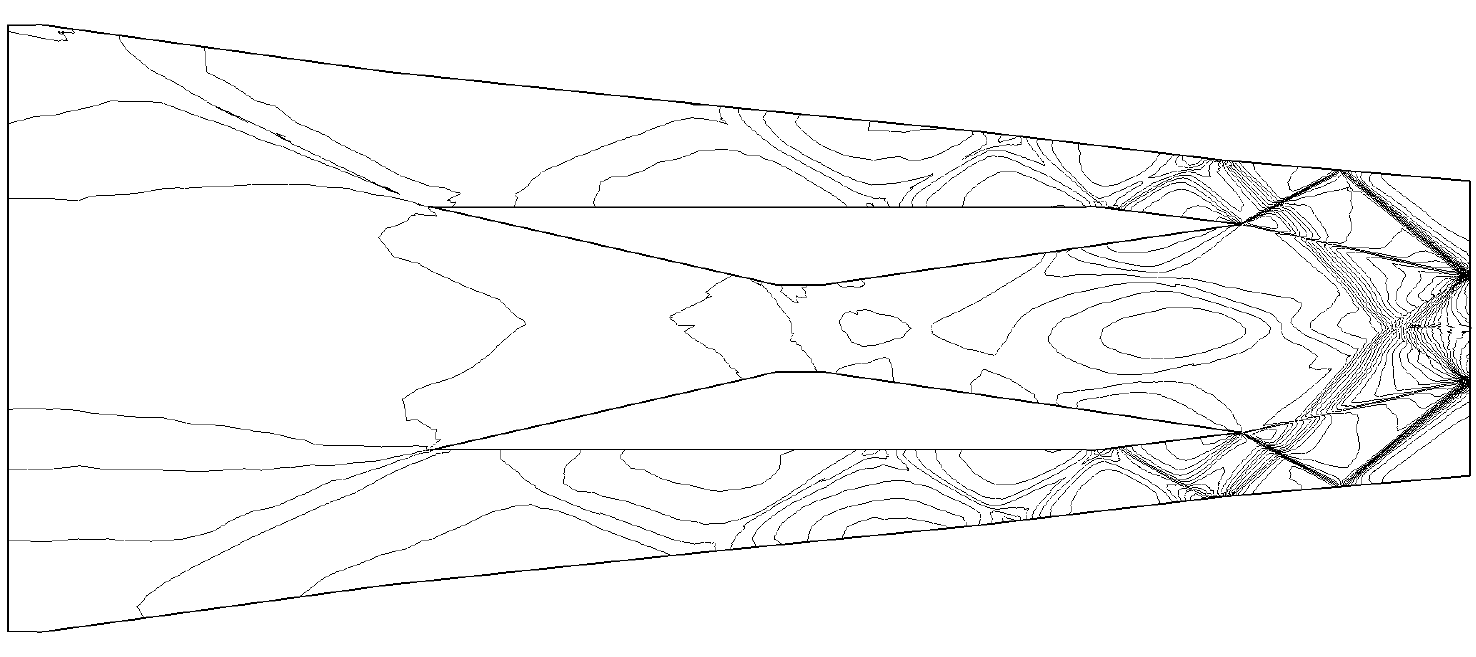}
\caption{\label{fig6}The scramjet: mesh (top), level lines of the density (middle) and of the adjoint density(bottom)}
\end{center}
\end{figure}

The flow contains many shocks and some contact discontinuity.  These emanate from the rear tip of nozzle and the adjoint density is clearly discontinuous there.  By contrast it is not discontinuous at the shock lines of the density.  

However shocks at the outflow boundary create discontinuous boundary conditions for the adjoints which, by analogy with Burgers' case,  will propagate as discontinuities from right to left in the adjoint. The discontinuities are not on the flow shocks but around them on the characteristics which ends at the intersection of the shocks and the outflow boundary as in the one dimensional example of section \ref{s4}. 

\paragraph{Boundary Conditions on the Outflow Boundary for the adjoint}
When the adjoint  is generated automatically by A.D  and hand transposition of the discrete equations we lose control over the boundary conditions at the outflow boundary, because are generated automatically and implement in weak form in the finite volume code.  

On figure \ref{fig5} the analytical values (\ref{adjcond}) are compared with the values generated by hand differentiation of the discrete solver.
They agree quite well and there was no noticeable difference when the full Jacobian of the flux function computed by A.D. was used instead of the first order approximation.

We noticed however that the shocks tended to be resolved better by the analytical expressions. Note also that $W^*_3$ has a shock but no Dirac singularity in theory while the computed value seem to have a spurious Dirac singularity.
\begin{figure}
\begin{center}
\includegraphics[width=6cm]{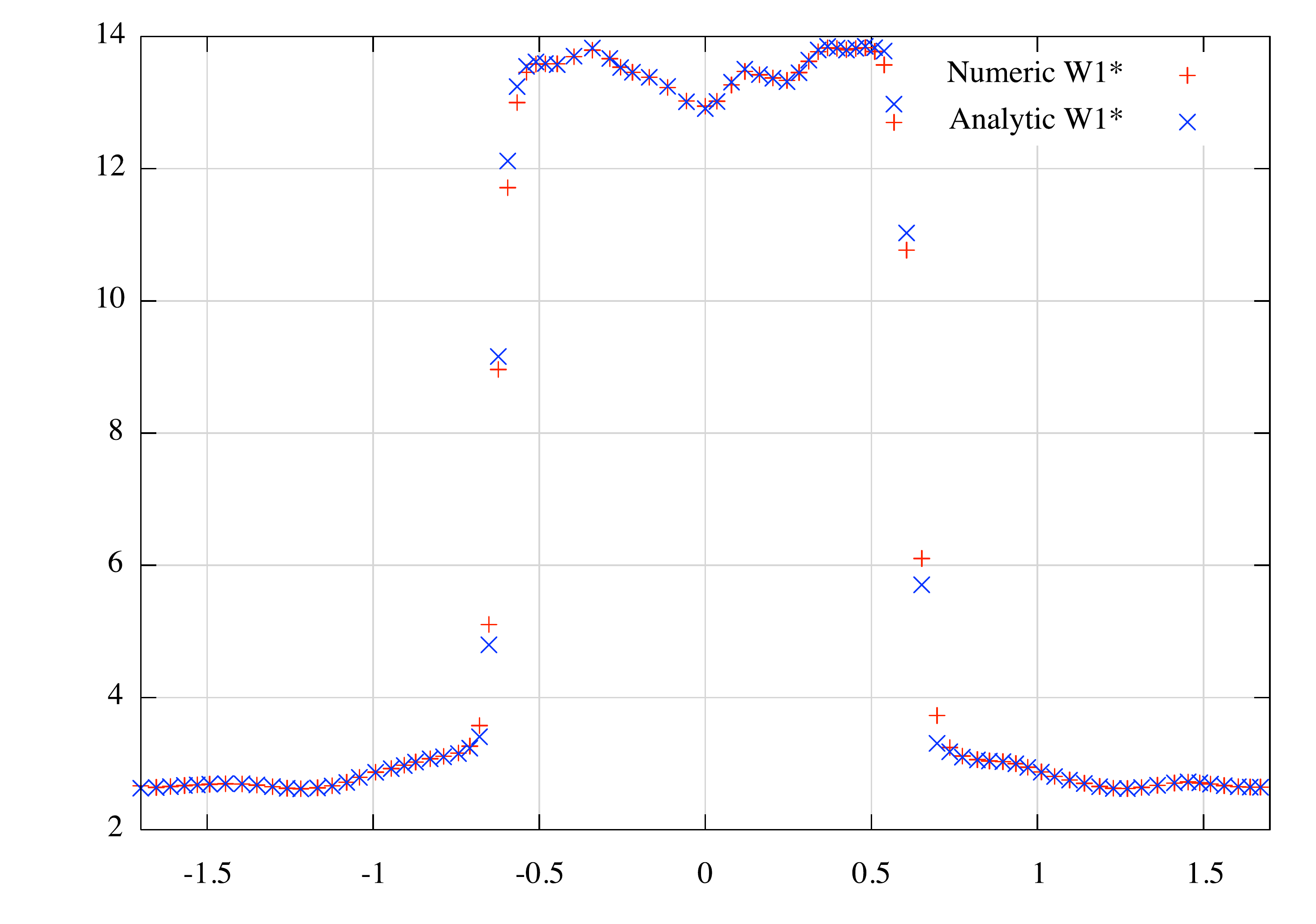}
\includegraphics[width=6cm]{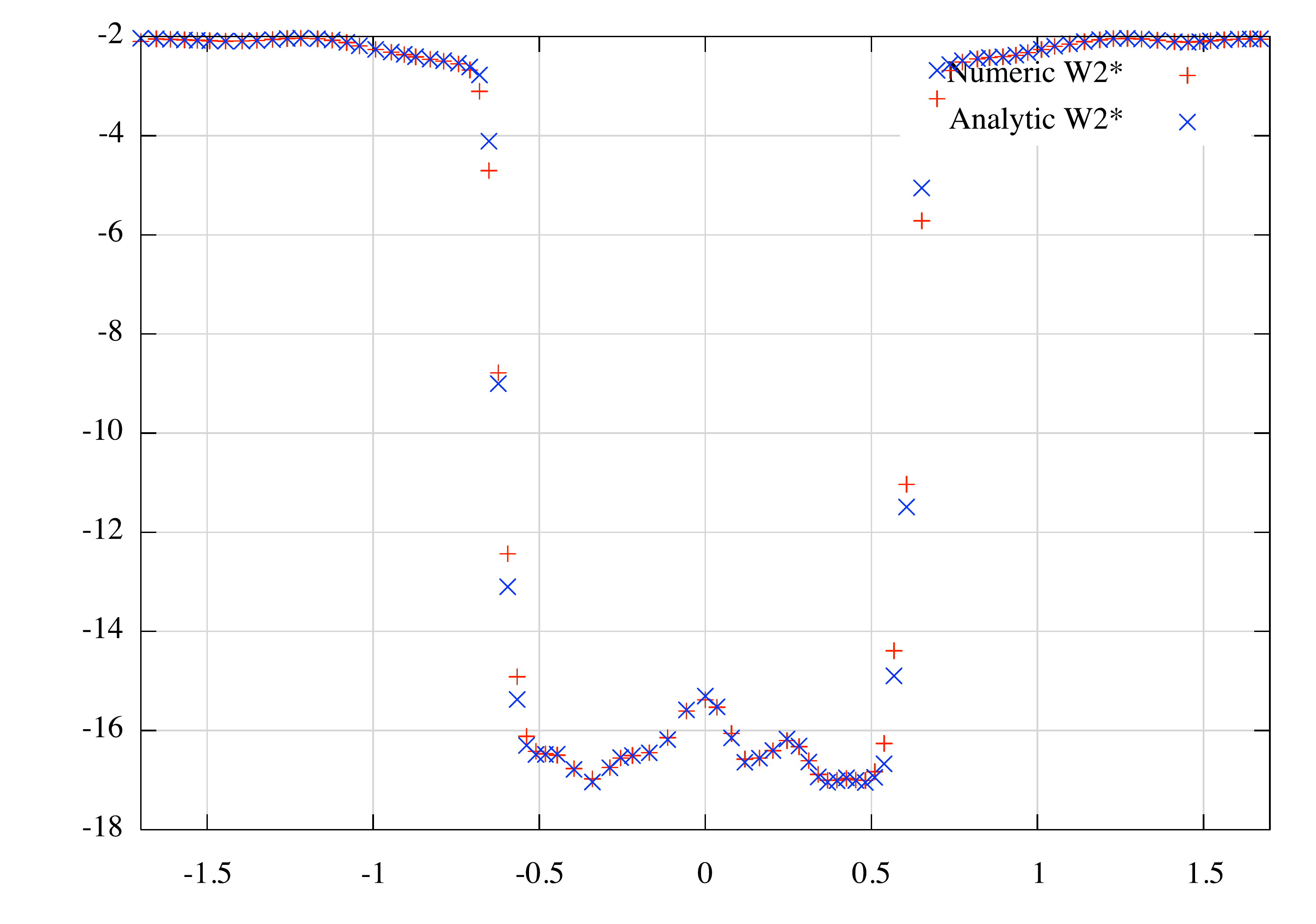}
\includegraphics[width=6cm]{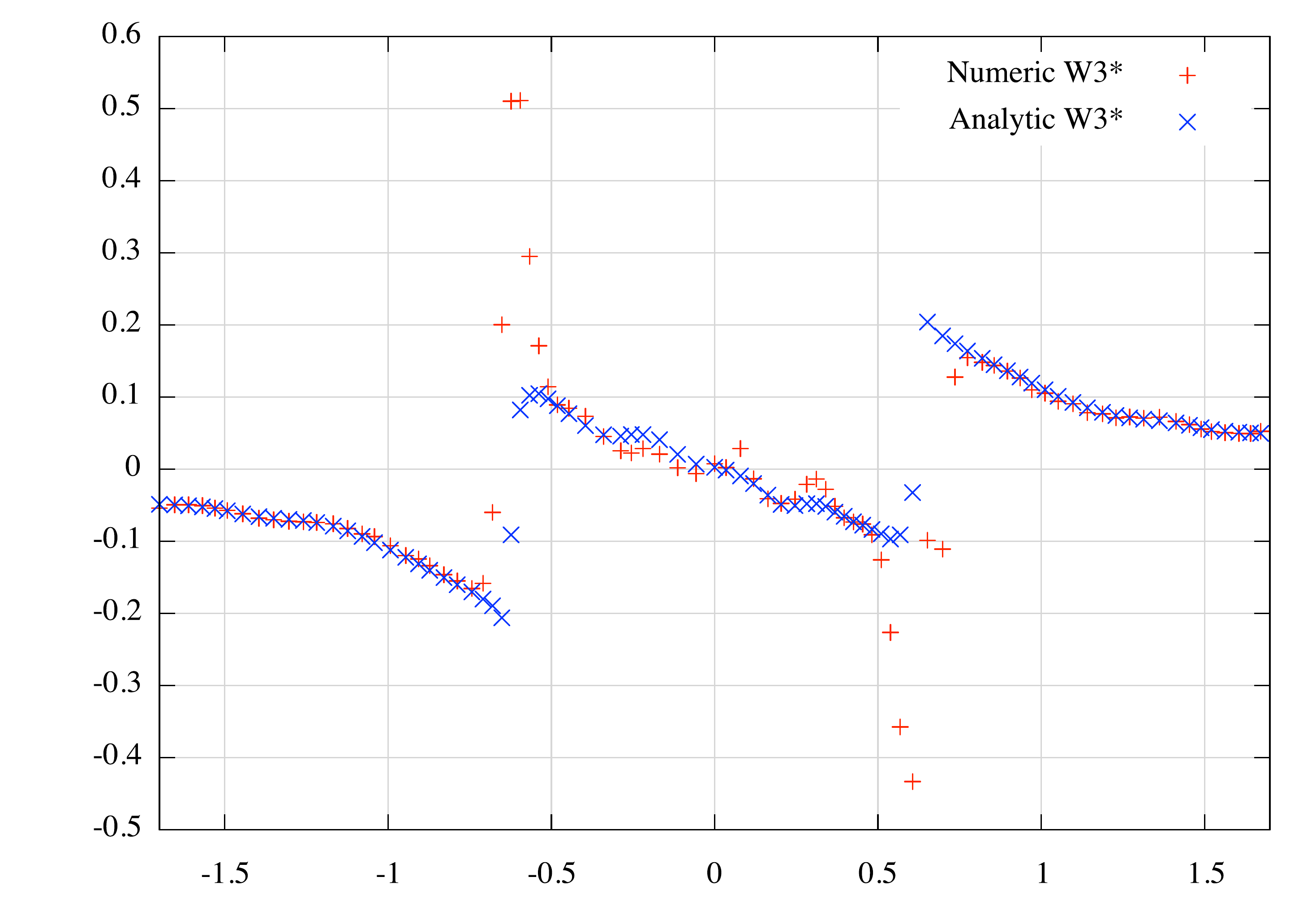}
\includegraphics[width=6cm]{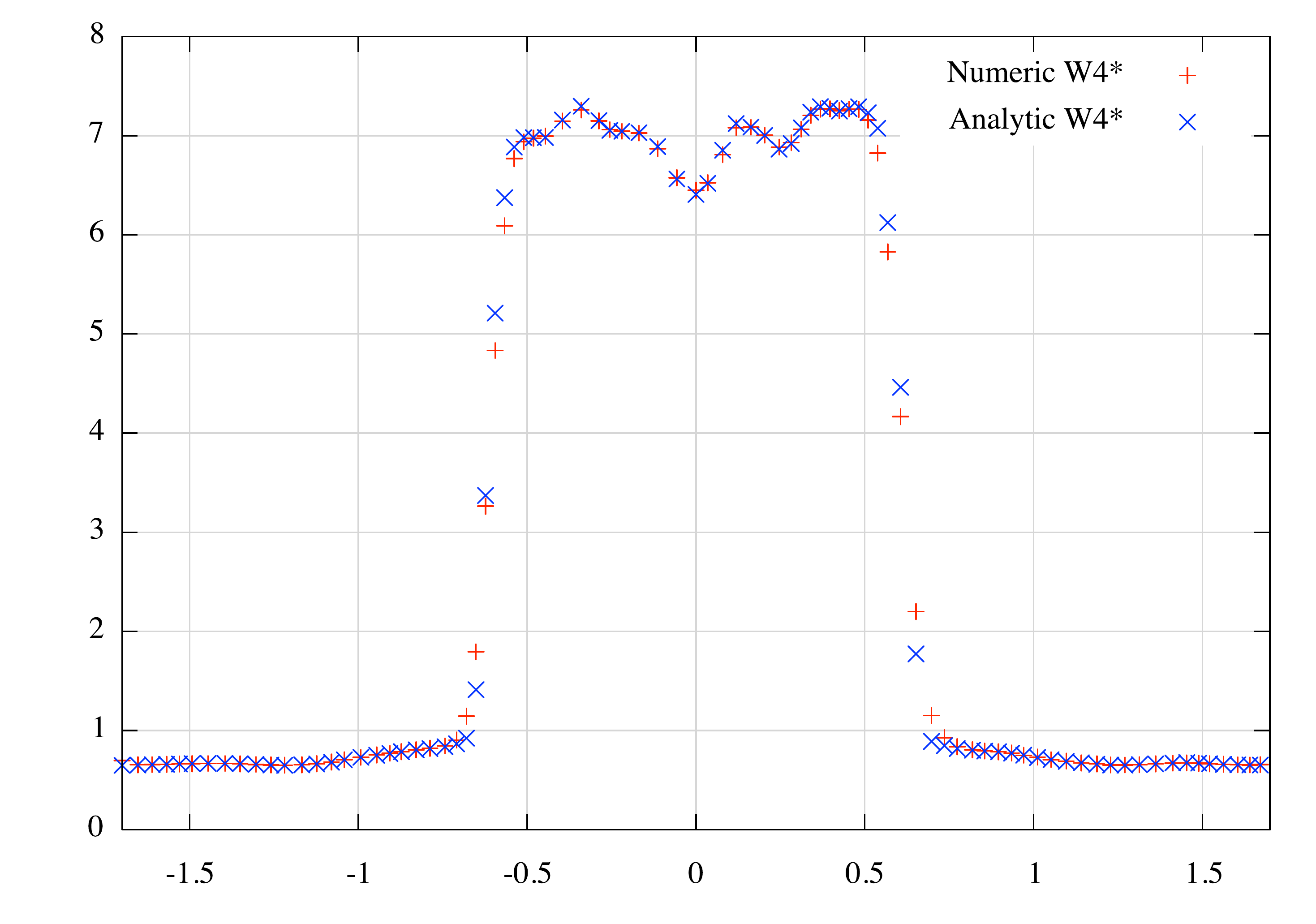}
\caption{\label{fig5}The scramjet, verification of (\ref{adjcond}).}
\end{center}
\end{figure}

\section{Optimal Wing Profile with least Sonic Boom}
Consider an airfoil $\Sigma$ at rest in a semi-infinite air domain $\Omega$.
Air comes in from the left boundary $L$ at supersonic speed and exits on the right by boundary $R$.
Let the ground $S$ be the lower horizontal part of the boundary ;
thus the boundary of the domain is $\p\Omega = \Sigma\cup L\cup S\cup R$. The control criteria is on the pressure $p$:
\[
J = \frac12\int_S(p-p_0)^2,\hbox{~~ ~therefore~~~}\delta J = \int_S(\overline{p}-\overline{p_0})\delta p.
\]
The aim is to find the best admissible $\Sigma$ to minimize $J$.

\subsection{Boundary Conditions}
\paragraph{Outflow Boundary Conditions on the Adjoint on $R$}
The flow is supersonic so the situation is essentially the same as in section \ref{sec31} for the scramjet except that $J$ has no part on $R$ so (\ref{50a}) has zero on the right hand side and both components of the normal $n$ appear:
\begin{eqnarray*}
W^*\cdot F'(W). \vec{n} = 0
\end{eqnarray*}
As $ F'(W). \vec{n}$ is a non-singular matrix when the flow is hypersonic, this equation implies that all four components of $W^*$ are zero on $R$:
\eq{\label{BConR}
W^*=0 \hbox{ on }R.
}
\paragraph{Inflow Boundary Conditions for the Adjoint on $L$}
The flow is supersonic on $L$ so all components of the linearized flow $\delta W$ are equal to zero.  All the characteristics are coming in so for the adjoint state the characteristics are coming out and hence
\eq{\label{bconl}
\hbox{$W^*$ needs no boundary condition on $L$.}
}

\paragraph{Boundary Conditions for the Adjoint on the Ground $S$}

The shock is assumed to reflect perfectly so the vertical component of the velocity is zero: $W_3=0$.

There is only one incoming characteristics for the linearized Euler system and for its adjoint as well.  Assume that $\delta W_3=0$ is the boundary condition for the linearized Euler system. In \cite{gilespierce}, Giles and Pierce gave a constructive argument to choose the boundary condition for the adjoint state, in this case $W^*_3$ should be given.  Let us verify directly that it is the right choice.

On $S$ the normal is $(0,1)^T$ and the velocity is $(u,0)^T$, so
{\small
\[
n\cdot F'(W)= \left( \begin{array}{cccc}
0 & 0 & 1 & 0 \\
0 & 0 & u & 0 \\
\displaystyle \frac{(\gamma-1)}{2} u^2  & -(\gamma-1)u & 0 & \gamma -1 \\
\displaystyle 0& 0 &
\displaystyle \gamma E - \frac{\gamma-1}{2} u^2 & 0
\end{array} \right)
\]
}
Therefore
{\small\eq{\label{50b}&&
W^*\cdot(n\cdot F'(W))\delta W
= \left(\matrix{0&0& \frac{(\gamma-1)}{2} u^2\delta W_1 -(\gamma-1)u\delta W_2 + (\gamma -1)\delta W_4&0}\right)W^*
\cr&&
= (\gamma-1)(\frac{u^2}{2} \delta W_1 -u\delta W_2 + \delta W_4)W^*_3
}}
On the other hand
\[
\displaystyle p = (\gamma-1) \rho e - \frac{\gamma -1}{2} \frac{(\rho u)^2 + (\rho v)^2}{\rho}=(\gamma-1)(W_4-\frac{W_2^2+W^2_3}{2 W_1}),
\]
therefore
\[
\displaystyle\delta p = (\gamma-1)(\frac{W_2^2}{2 W_1^2}\delta W_1
-\frac{W_2}{W_1}\delta W_2 + \delta W_4)
= (\gamma-1)(\frac{u^2}2\delta W_1 - u\delta W_2 + \delta W_4).
\]
By choosing 
\[
W^*_3|_S = \overline{p-p_0},
\]
   (\ref{50})  becomes $(p-p_0)\delta p$ and in the stationary regime (\ref{dual}) becomes
\eq{\label{w3}
	\delta J = -\int_{\p\Omega\backslash S}W^*\cdot(n\cdot\overline{F'(W)}\delta W)
}

\paragraph{Boundary Conditions for the Adjoint on the airfoil $\Sigma$}
The flow is subsonic and tangent to $\Sigma$ so $u\cdot n=0$. As for $S$ one boundary condition is required for $\delta W$. Let $\delta W\cdot n$ be given. An elementary calculus gives
\eq{
W^*\cdot (F'(W)\cdot n)\delta W
&=& (\gamma-1)W^*\cdot n (\frac12(u^2+v^2)\delta W_1 -u\delta W_2 - v \delta W_3 + \delta W_4)
\cr&&
+(W^*_1+u W^*_2 + v W^*_3)\delta W\cdot n
}
As Giles and Pierce found \cite{gilespierce} by another method, this expression contains no other $\delta W$-term but $\delta W\cdot n$ if and only if
\eq{\label{bconsig}
W^*\cdot n = 0 \hbox{ on }\Sigma.
}
\begin{proposition}
Let $W^*$ be defined by
\eq{\label{prop}&&
\p_T W^* + \overline{F'(W)}^T\n W^* = 0,~~~W(T) = 0,
\cr&&
W^*\cdot n = 0 \hbox{ on }\Sigma~~~
\cr&&
W^*|_R=0,~~~W^*_3|_S = \overline{p-p_0}
}
Then, asymptotically in time,
\eq{\label{magic}
\delta J = -\int_\Sigma(W^*_1+\vec U\cdot\vec W^*_{2.3} )\delta \vec W_{2.3}\cdot\vec n
}
where $W_{2.3}$ is the vector $(W_2,W_3)^T$.
\end{proposition}
Formula (\ref{magic}) is essential to set up descent algorithms for optimal shape design to minimize $J$.  The proof of the proposition is a consequence of (\ref{dual}),(\ref{bconl})(\ref{w3}),(\ref{bconsig}). We have assumed that there was no shock on $\Sigma$; if it is not so, mean values have to be used for the integrand in (\ref{magic}).

\begin{remark}
It may be more readable to rewrite (\ref{magic}) in terms of the components of the adjoint state $(\rho^*,(\rho U)^*, (\rho e)^*)$:
\eq{\label{osde}
\delta J =  -\int_\Sigma(\rho^* + \vec U\cdot\vec{(\rho U)^*}) \delta(\rho \vec U)\cdot\vec n
}

\end{remark}

\subsection{A Gradient Method for Shape Optimization}

Thanks to (\ref{osde}) we have now a method to modify the shape of the airfoil $\Sigma$ to decrease $J$.

Following \cite{op,bmop}, consider a variation $\Sigma_\alpha$ of $\Sigma$ defined by the normal displacement $\alpha$ and given by
\[
\Sigma_\alpha = \{ x+\alpha(x)n(x)~:~ x\in\Sigma\}
\]
Denote $W_\alpha$ the new flow variables and to simplify notations introduce $V:=\rho U$  By definition we have
\[
V(x)\cdot n(x)=0,~~~V_\alpha(x+\alpha n(x))\cdot n_{\Sigma_\alpha}(x+\alpha n(x))=0,~~~\forall x\in\Sigma
\]

Assuming differentiability and ignoring higher order terms, the second equation implies
\begin{equation}
\displaystyle{
0= V_\alpha(x)\cdot n(x) + \alpha \left(~\left(\n V(x)\cdot n(x)\right)\cdot n(x) + V(x)\cdot n'(x)~\right)
}
\end{equation}
where $n'$ is the derivative with respect to $\alpha(x)$ of $n_{\Sigma_\alpha}$.
The above is also
\[
\delta V\cdot n|_\Sigma = V_\alpha(x)\cdot n(x) = -\alpha(\frac{\p V_n}{\p n} - \kappa V_t)
\]
where $t$ is the tangent vector associated with $n$ -- i.e. $n$ rotated clockwise by $90$ degrees -- because $n'=-\kappa t$ where $\kappa$ is the curvature (inverse of the radius of curvature).

Thus (see (\ref{osde}) ) by choosing
\[
\alpha = -\lambda(\rho^* + \vec U\cdot\vec{(\rho U)^*})(\frac{\p (\rho U_n)}{\p n} - \kappa \rho U_t)
\]
for a small enough constant scalar $\lambda$, $J$ will decrease because, in this case
\[
\delta J = -\lambda\int_\Sigma (\rho^* + \vec U\cdot\vec{(\rho U)^*})^2(\frac{\p (\rho U_n)}{\p n} - \kappa \rho U_t)^2 + o(\lambda)
\]
This method was used by B. Mohammadi in \cite{bmop} and G. Rog\'e et al in \cite{roge} to reduce the sonic boom of a business jet by 30\% without affecting the lift.  Automatic differentiation was used, so we intend to reproduce the results with the code presented here in the future.  Meanwhile, as for the scramjet, let us verify that the discrete adjoint converges to the continuous one.

\subsection{Numerical Tests}
A supersonic flow at Mach 2 is computed around a NACA0012 airfoil flying over ground at very low altitude.
(see figure \ref{fig7}).  Euler flow and its discrete adjoint are computed on a flow-adapted mesh in 2D.  The value of the adjoint on the ground $S$ is compared with its analytical value.

Results shown on figure \ref{fig8} confirm that the discrete adjoint converges to the continuous one and that it is continuous at the shocks.

\begin{figure}
\begin{center}
\includegraphics[width=6cm]{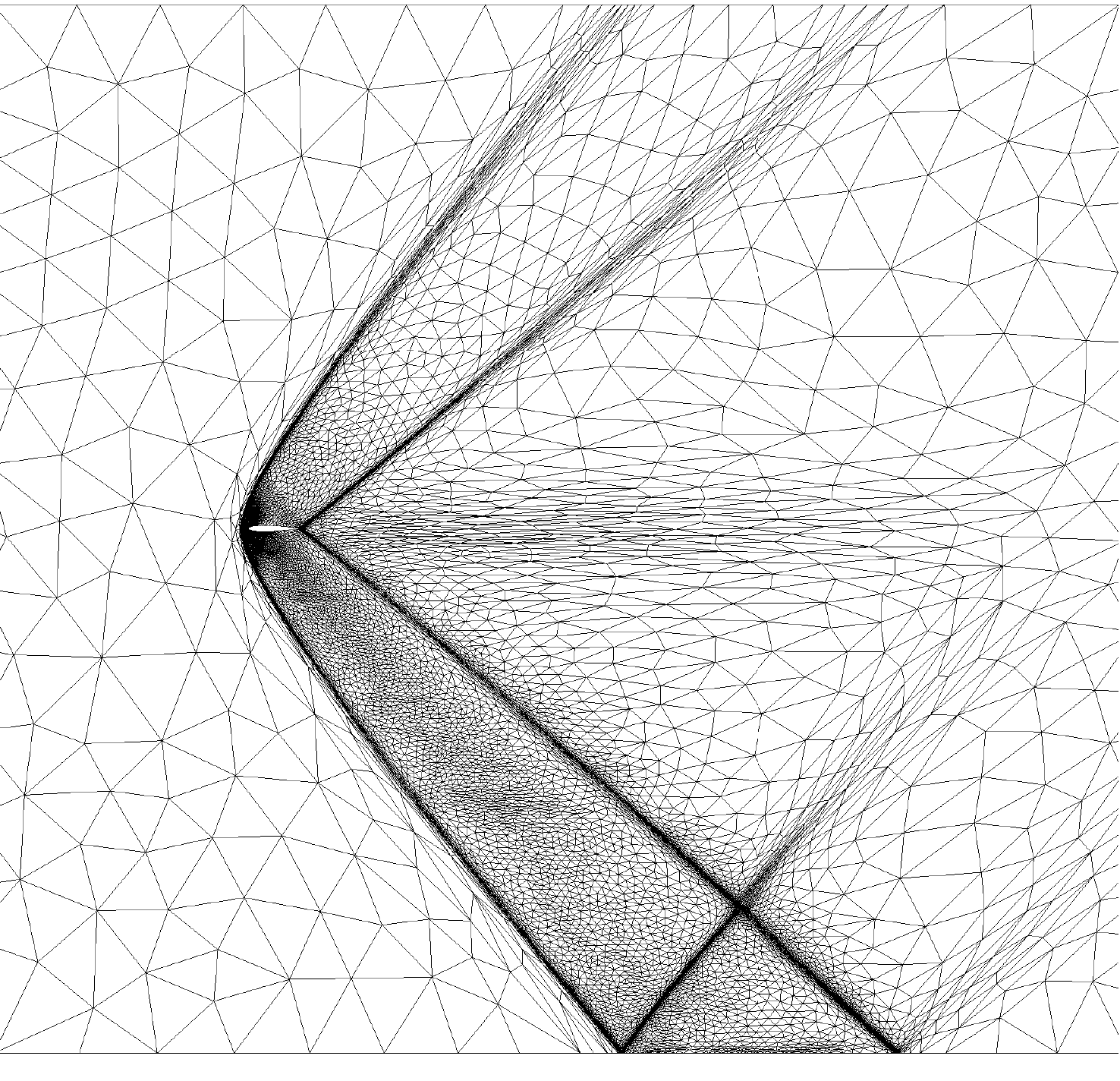}
\includegraphics[width=6cm]{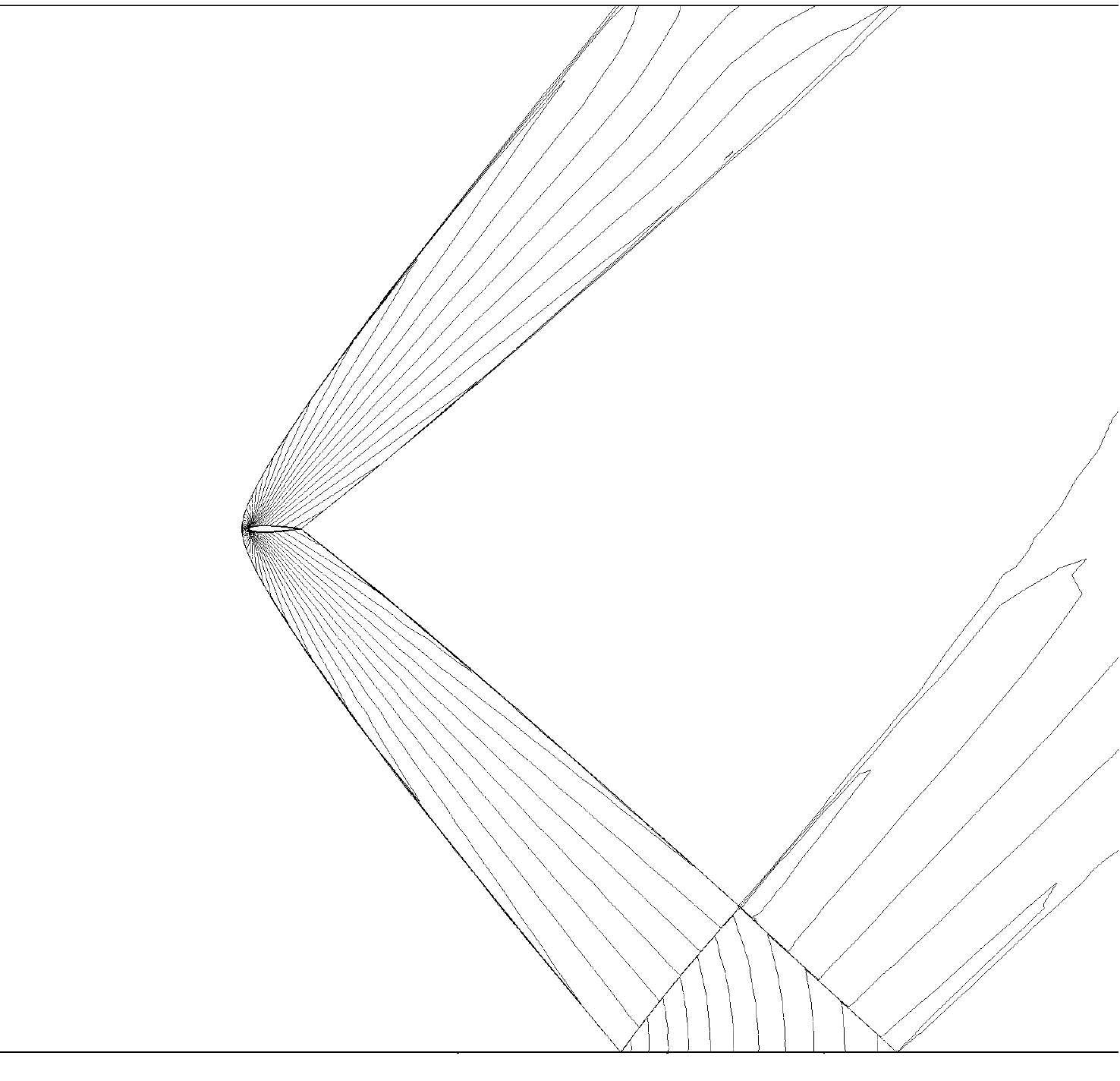}\\
\includegraphics[width=12cm, height=6cm]{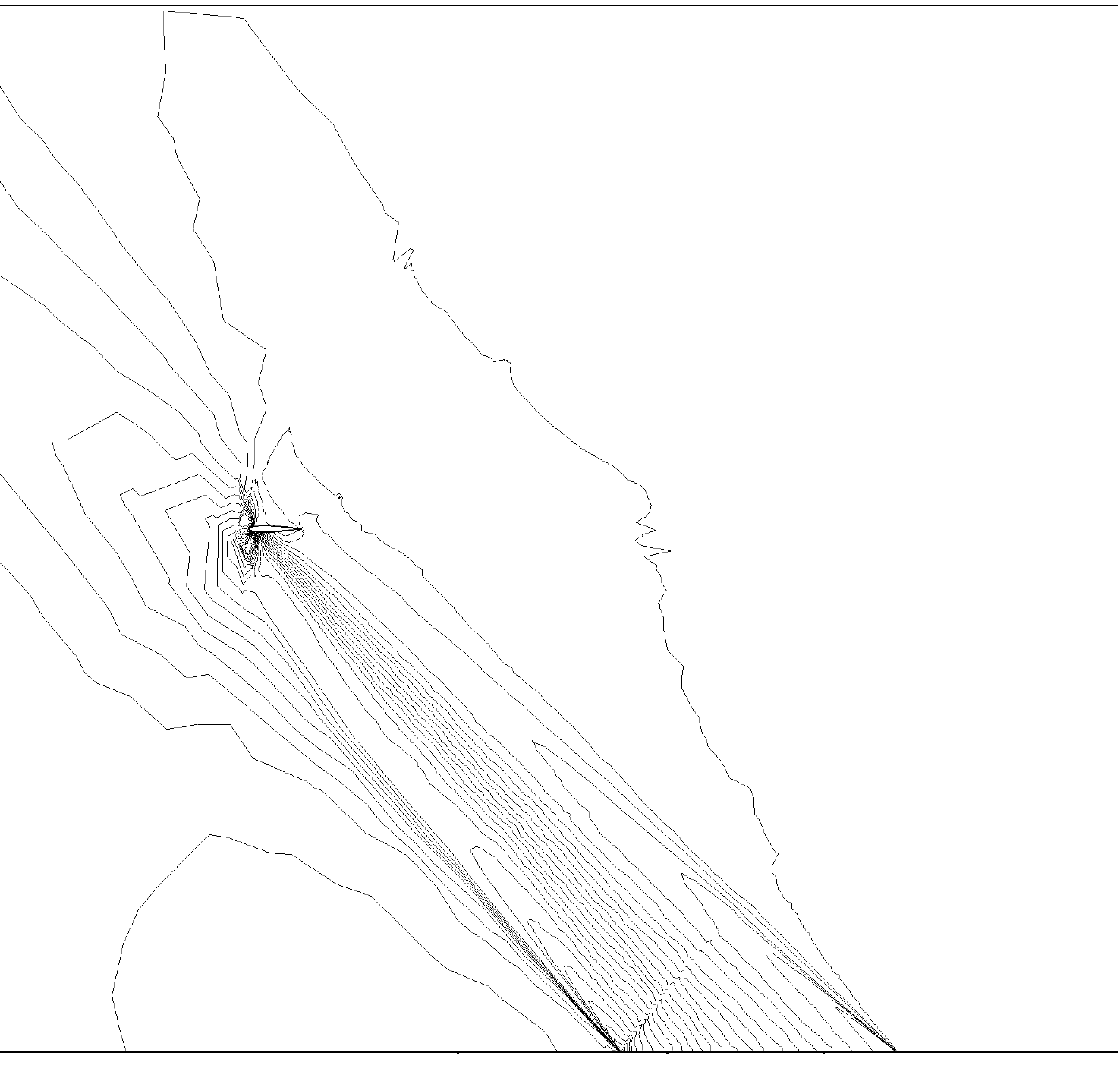}
\caption{\label{fig7} The NACA airfoil: the adapted mesh (top left),; the level lines of the density (top right) and (bottom, the figure is stretched horizontally) the level lines of the first component of the adjoint, i.e. the adjoint density.}
\end{center}
\end{figure}

\paragraph{Analysis of Figure \ref{fig7}}
Here too we observe that the adjoint density is continuous at the shocks of $\rho$ except near the ground.  There the boundary condition on the adjoint forces a discontinuity which, by analogy with Burgers', should spread into shocks on the characteristics at the shock point.
Since there is only one incoming characteristic we expect to have only one shock in the adjoint (while there was 2 in the case of Burgers').  On the reflected shock near the ground, the adjoint has a bump type variation. It is not clear whether it is physical or numerical.  The adjoint also varies greatly near the NACA profile. A zoom shows that these variations are not discontinuities.

\paragraph{Discrete versus Continuous Adjoint (Figure \ref{fig8})}
For the minimizing of the sonic boom we have seen that the theory predicts that the boundary condition of the third component of the adjoint $W_3^*$ is $p-p_0$.  On Figure \ref{fig8} both quantities are displayed; the concordance is excellent. This is another indication that the adjoint obtained by automatic differentiation in reverse mode of the discrete solver converges to the continuous adjoint when the mesh is refined.

\begin{figure}
\begin{center}
\includegraphics[width=10cm]{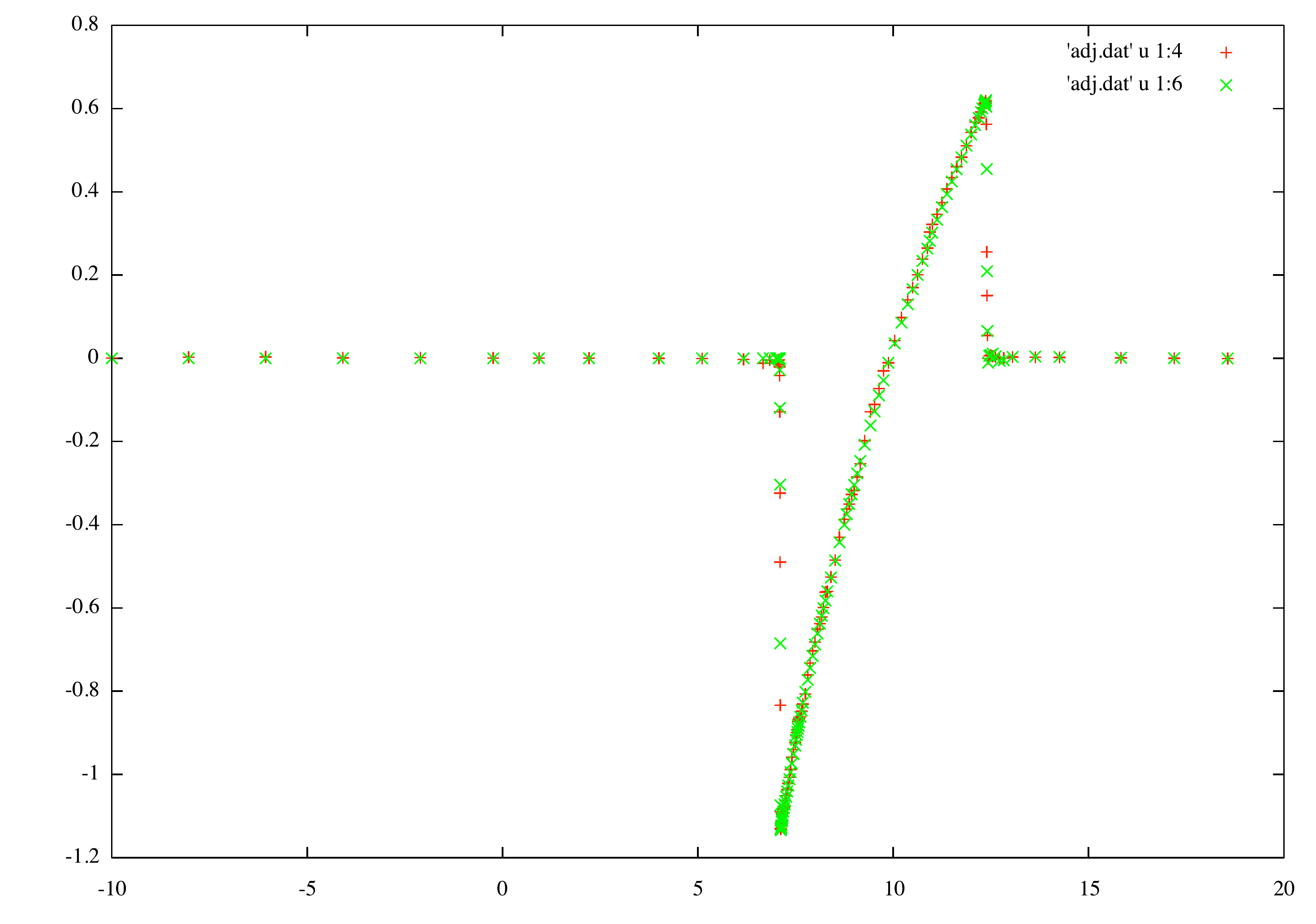}
\caption{\label{fig8}The NACA airfoil:  comparison between $W_3^*$, the component of the adjoint in duality with $\rho v$ and $p-p_0$ on $S$. Although the adjoint's equation is generated by automatic differentiation for the discrete case, it agrees with the continuous limit (see (\ref{prop}))}
\end{center}
\end{figure}

\section*{Conclusion}
Industry is never happy to include an unreadable portion of code in their numerical tools; this is one reason to avoid using automatic differentiation to generate the adjoint of the Euler linearize system.  By contrast A.D. is very practical and avoids code errors but theory doe not support yet the convergence when the mesh size decreases.

On the whole this numerical study validates both the automatic generation of adjoints and the hand derivation with incomplete derivatives for the flux functions (i.e. without differentiation of the MUSCL and slope limiters).  It is also reassuring to see on two examples that the discrete adjoints seem to converge to the   continuous adjoints.

One difficulty remains unanswered: whether the automatically generated adjoints are correct in the case of discontinuous boundary data.  The theory requires a precise boundary condition at the shock which the discrete adjoints don't seem to observe and even if they did a rather fine mesh would be needed to implement the condition anyway. However on the scramjet and on the airfoil this difficult did not show up in the numerical tests.

\newpage

\section{Appendix A: Calculus of Variation for Discrete Incompressible Euler Flows}
With reference to (\ref{pbe}) and (\ref{p1p2}),
\eq{&&
\delta J_h = \int_{D\times(0,T)}(u_h^{m+1}-u_d)\delta u^{m+1} \hbox{ with}
\cr&&
	\int_\Omega(\frac{\delta u_h^{m+1}}{\delta t}+\n \delta p^{m+1})v_h - \int_\Omega q_h\n\cdot \delta u_h^{m+1} 
\cr&&
= \frac1{\delta t}\int_\Omega{v_h(x)}[\delta u_h^m(y)-\delta t \delta u_h^m(x)\cdot \n u_h^m(y)]|_{y=x-\delta t u_h^m(x)} \d x
}
Now with a change of variable $y=x-\delta t u_h^m(x)$ in the first half of the last integral, it is also
\eq{
\frac1{\delta t}\int_{\Omega^+} v_h(y+\delta t u_h^m(x)) \delta u_h^m(y)\d y -
\int_\Omega v_h(x)\delta u_h^m(x)\cdot \n u_h^m(x-\delta t u_h^m(x)) \d x
}
where $\Omega^+=\{y:=x-\delta t u_h^m(x): x\in\Omega\}$.

The adjoint equation is obtained by replacing $\delta u_h,\delta p_h$ by $ v_h,q_h$ and $v_h,q_h$ by $ u_h^*,p_h^*$ and adding the contribution of $\delta J$ at time $(m+1)\delta t$ on the right hand side:
\eq{&&
	\int_\Omega(\frac{v_h}{\delta t}+\n q_h){u^*_h}^m - \int_\Omega {p^*_h}^m\n\cdot v_h=\int_D (u_h^{m+1}-u_d) v_h
	\cr&&
+ \frac1{\delta t}\int_{\Omega^+} {{u^*_h}^{m+1}(y+\delta t u_h^{m+1}(x))}v_h(y)\d y 
-\int_\Omega v_h {u_h^*}^{m+1}\cdot \n {u^{m+1}_h}(y) \d x~~~~~  
}
for all $v_h,q_h$ with $v_h|_{\Gamma^+}=0$, $v_h\cdot n|_{\Gamma}=0$.

We choose to impose $u_h^*(T)=0$, $u^*_h|_{\Gamma^+}=0$, $u^*_h\cdot n|_{\Gamma}=0$.

Let us replace $ (v_h,q_h)$ by $(\tilde\delta u_h,\delta p_h)$ where the tilde means that the value of $\delta u_h^{m+1}$ at the boundary nodes of $\Gamma^-$ have been put to zero to be admissible:
{\small
\eq{&&
\delta J =	\sum_m\delta t[ \int_\Omega(\frac{\tilde\delta u_h^{m+1}}{\delta t}+\n\delta p_h^{m+1}){u^*_h}^{m} - \int_\Omega {p^*_h}^{m}\n\cdot \tilde\delta u_h^{m+1}] - I \hbox{ with }
\cr&&
I= \sum_m[\int_{\Omega^+} {{u^*_h}^{m+1}(x)}\tilde\delta u_h^{m+1}(y)\d y 
-\delta t\int_\Omega \tilde\delta u_h^{m+1}(x) {u_h^*}^{m+1}(x)\cdot \n {u^{m+1}_h}(y) \d x]
\cr&&
=  \sum_m[\int_{\Omega} {{u^*_h}^{m}(x)}\tilde\delta u_h^{m}(x-\delta t u_h^{m}(x))\d x 
-\delta t\int_\Omega \tilde\delta u_h^{m}(x) {u_h^*}^{m}(x)\cdot \n {u^{m}_h}(y)\d x]~~ 
}}
It contains the equation of $(\delta u_h,\delta p_h)$ with the test functions replaced by $(u_h^*,p_h^*)$ it  if it wasn't for the tildes.  Let us call $\delta a_h$ the vector value $P^2$ finite element function which is equal to $\delta a$ at the nodes of $\Gamma_h^-$ and zero at other nodes.  Since $\delta u_h=\tilde\delta u_h + \delta a_h$ at all time steps, taking the boundary conditions into account we have up to $O(\delta t),o(\delta a_h)$ terms:
\eq{\label{dgrad}
\delta J = \delta t\sum_m\int_\Omega({u_h^*}^m \cdot(u^m_h\cdot\n\delta a_h)+ \delta a_h\cdot({u_h^*}^m\cdot\n u_h^m)
-{p_h^*}^m\n\cdot\delta a_h ).
}

\section*{Appendix B: Jacobian matrix in 2D}\label{JacobianMatrix}

In two dimensions, the Euler equations could be symbolically rewritten:
\begin{equation}
\displaystyle{\partial_t W}+ \nabla \cdot F(W) = 0 \,,
\end{equation}
where $W=(\rho,\rho u,\rho v, \rho E)^T$ is the conservative variables vector
and the vector $F$ represents the convective operator.
The vector $F$ may be decomposed as $F(W) = F_1(W) \,e_x + F_2(W) \,e_y$ with
$\displaystyle p = (\gamma-1) \rho E - \frac{\gamma -1}{2} \frac{(\rho u)^2 + (\rho v)^2}{\rho}$ and
$$
F_1(W) = \left( \begin{array}{c}
\rho u \\ \rho u^2 + p \\ \rho u v \\ (\rho E + p) u
\end{array} \right) \; \mbox{and} \; \;
F_2(W) = \left( \begin{array}{c}
\rho v \\ \rho u v \\ \rho v^2 + p \\  (\rho E + p) v
\end{array} \right)
$$
The Jacobian of $p$ with respect to the conservative variable is:
\begin{equation}\label{JacP}
\frac{\partial p(W)}{\partial W} = (\gamma-1) \,
{}^t \left( \frac{q^2}{2} \,,\, -u \,,\, -v \,,\, 1  \right)
\end{equation}
where $q^2=u^2+v^2$.
After simplifications, the Jacobians read:
{\small
$$
\frac{\partial F_1(W)}{\partial W} = \left( \begin{array}{cccc}
0 & 1 & 0 & 0 \\
\displaystyle \frac{(\gamma-1)}{2} q^2 - u^2 & -(\gamma-3)u & -(\gamma-1) v & \gamma -1 \\
-uv & v & u & 0 \\
\displaystyle ((\gamma -1) q^2 - \gamma E) u &
\displaystyle \gamma E - \frac{\gamma-1}{2} \left( 2 u^2+q^2 \right) & -(\gamma-1)uv & \gamma u
\end{array} \right) ,
$$

$$
\frac{\partial F_2(W)}{\partial W} = \left( \begin{array}{cccc}
0 & 0 & 1 & 0 \\
-uv & v & u & 0 \\
\displaystyle \frac{(\gamma-1)}{2} q^2 - v^2 & -(\gamma-1)u & -(\gamma-3) v & \gamma -1 \\
\displaystyle ((\gamma -1) q^2 - \gamma E) v & -(\gamma-1)uv &
\displaystyle \gamma E - \frac{\gamma-1}{2} \left( 2 v^2+q^2 \right) & \gamma v
\end{array} \right) ,
$$
}
For an edge $\bf{e}$ with $\bf{n}$ as normal vector, we have:
$\displaystyle A(W)  :=  \frac{\partial F(W). \bf{n}}{\partial W}=$
{\small
\begin{eqnarray*}\left( \begin{array}{cccc}
0 & n_x & n_y & 0 \cr
\displaystyle \frac{\gamma_1}{2} q^2 n_x - u \, {\bf u \cdot n} & {\bf u \cdot n} - (\gamma-2)u n_x & u\,n_y - (\gamma-1) v n_x & \gamma_1 n_x \cr
\displaystyle \frac{\gamma_1}{2} q^2 n_y - v \, {\bf u \cdot n} & v\,n_x - \gamma_1 u n_y & {\bf u \cdot n} - (\gamma-2)v n_y & \gamma_1 n_y \cr
\displaystyle \left( \gamma_1 q^2 - \gamma E \right) {\bf u \cdot n}  &
\displaystyle \left( \frac{p}{\rho} + E \right) n_x - \gamma_1 u {\bf u \cdot n} &
\displaystyle \left( \frac{p}{\rho} + E \right) n_y - \gamma_1 v {\bf u \cdot n} & \gamma {\bf u \cdot n}
\end{array} \right)
\end{eqnarray*}
}
where ${\bf u} = (u,v)$, $\gamma_1=\gamma-1$ and ${\bf n}=(n_x,n_y)$.

\end{document}